\documentclass[useAMS,usenatbib,a4paper]{mn2e}
\usepackage{graphicx}
\usepackage{amsmath}
\usepackage{amssymb}

\def\ucb{1}
\def\brera{2}
\def\ioa{3}
\def\lanl{4}
\def\steward{5}
\def\clemson{6}
\def\mpig{7}
\def\univ{8}
\def\inaf{9}
\def\ifa{10}
\def\iac{11}
\def\iap{12}
\def\eso{13}
\def\ucsc{14}

\title[Evidence of SN Dust from GRB 071025] {Evidence for Supernova-Synthesised Dust from the Rising Afterglow of GRB 071025 at z$\sim$5}

\author[D. A. Perley et al.]
{Daniel A. Perley$^{\ucb}$\thanks{e-mail: dperley@astro.berkeley.edu}, 
J.~S.~Bloom$^{\ucb}$, 
C.~R.~Klein$^{\ucb}$, 
S.~Covino$^{\brera}$,
T.~Minezaki$^{\ioa}$, 
\newauthor 
P.~Wo{\'z}niak$^{\lanl}$, 
W.~T.~Vestrand$^{\lanl}$, 
G.~G.~Williams$^{\steward}$, 
P.~Milne$^{\steward}$
N.~R.~Butler$^{\ucb}$,
\newauthor 
A.~C.~Updike$^{\clemson}$, 
T.~Kr\"uhler$^{\mpig,\univ}$, 
P.~Afonso$^{\mpig}$, 
A.~Antonelli$^{\inaf}$,
L.~Cowie$^{\ifa}$,
\newauthor 
P.~Ferrero$^{\iac}$, 
J.~Greiner$^{\mpig}$, 
D.~H.~Hartmann$^{\clemson}$, 
Y.~Kakazu$^{\iap}$, 
A.~K\"upc\"u~Yolda\c{s}$^{\eso}$,
\newauthor 
A.~N.~Morgan$^{\ucb}$, 
P.~A.~Price$^{\ifa}$,
J.~X.~Prochaska$^{\ucsc}$, 
Y.~Yoshii$^{\ioa}$\\
\\
$^{\ucb}${Department of Astronomy, University of California, Berkeley, CA 94720-3411, USA.} \\
$^{\brera}${INAF/Osservatorio Astronomico di Brera, via Bianchi 46, 23807 Merate, LC, Italy} \\
$^{\ioa}${Institute of Astronomy, School of Science, University of Tokyo, Mitaka, Tokyo 181-0015, Japan} \\
$^{\lanl}${Los Alamos National Laboratory, Los Alamos, NM 87545, USA} \\
$^{\steward}${Steward Observatory, University of Arizona, 933 North Cherry Avenue, Tucson, AZ 85721, USA} \\
$^{\clemson}${Department of Physics and Astronomy, Clemson University, Clemson, SC 29634-0978, USA} \\
$^{\mpig}${Max-Planck-Institut f\"ur extraterrestrische Physik, Giessenbachstra\ss e, D-85748 Garching, Germany} \\
$^{\univ}${Universe Cluster, Technische Universit\"{a}t M\"{u}nchen, Boltzmannstra\ss e 2, D-85748, Garching, Germany} \\
$^{\inaf}${INAF/Rome Astronomical Observatory, Via Frascati 33, 00044 Monte Porzio (Roma), Italy} \\
$^{\ifa}${Institute for Astronomy, 2680 Woodlawn Drive, Honolulu, Hawaii 96822-1839, USA} \\
$^{\iac}${Instituto de Astrof\'{\i}sica de Canarias, V\'{\i}a L\'{a}ctea s/n, 38200 La Laguna, Tenerife, Spain} \\
$^{\iap}${Institut d'Astrophysique de Paris, Paris, France} \\
$^{\eso}${European Southern Observatory, Karl-Schwarzschild-str. 2, 85748, Garching, Germany} \\
$^{\ucsc}${Department of Astronomy and Astrophysics, UCO/Lick Observatory; University of California, 1156 High Street, Santa Cruz, CA 95064, USA} \\
}

\begin{document}

\date{}

\pagerange{\pageref{firstpage}--\pageref{lastpage}} \pubyear{2009}

\def\LaTeX{L\kern-.36em\raise.3ex\hbox{a}\kern-.15em
    T\kern-.1667em\lower.7ex\hbox{E}\kern-.125emX}

\newcommand{\citepeg}[1]{\citep[{e.g.,}][]{#1}}
\newcommand{\citepcf}[1]{\citep[{see}\phantom{}][]{#1}}
\newcommand{\rha}[0]{\rightarrow}
\def\etal{{\sl et al.}}
\def\lsim{\hbox{ \rlap{\raise 0.425ex\hbox{$<$}}\lower 0.65ex\hbox{$\sim$}}}
\def\gsim{\hbox{ \rlap{\raise 0.425ex\hbox{$>$}}\lower 0.65ex\hbox{$\sim$}}}
\def\arcmin{\hbox{$^\prime$}}
\def\arcsec{\hbox{$^{\prime\prime}$}}
\def\arcdeg{\mbox{$^\circ$}}
\def\fd{\hbox{$~\!\!^{\rm d}$}}
\def\fh{\hbox{$~\!\!^{\rm h}$}}
\def\fm{\hbox{$~\!\!^{\rm m}$}}
\def\fs{\hbox{$~\!\!^{\rm s}$}}
\def\ale{\mathrel{\hbox{\rlap{\hbox{\lower4pt\hbox{$\sim$}}}\hbox{$<$}}}}
\def\age{\mathrel{\hbox{\rlap{\hbox{\lower4pt\hbox{$\sim$}}}\hbox{$>$}}}}
\def\msyr{\hbox{M$_\odot$ yr$^{-1}$}}
\def\Swift{{\textit{Swift}}\,}
\def\Fermi{{\textit{Fermi}}\,}
\def\apjl{{ApJL}\,}
\def\apj{{ApJ}\,}
\def\nat{{Nature}\,}
\def\aap{{A\&A}\,}
\def\pasp{{PASP}\,}
\def\aj{{AJ}\,}
\def\procspie{{Proc.~SPIE}\,}
\def\mnras{{MNRAS}\,}
\def\apjs{{ApJS}\,}
\def\aaps{{A\&AS}\,}
\def\araa{{ARAA}\,}
\def\apss{{Ap\&SS}\,}

\label{firstpage}
\label{lastpage}

\maketitle

\begin{abstract}
We present observations and analysis of the broadband afterglow of \Swift GRB~071025.  Using optical and infrared ($RIYJHK$) photometry, we derive a photometric redshift of $4.4 < z < 5.2$; at this redshift our simultaneous multicolour observations begin at $\sim$30~s after the GRB trigger in the host frame, during the initial rising phase of the afterglow.  We associate the light curve peak at $\sim$580~s in the observer frame with the formation of the forward shock, giving an estimate of the initial Lorentz factor $\Gamma_0 \sim 200$.  The red spectral energy distribution (even in regions not affected by the Lyman-$\alpha$ break) provides secure evidence of a large dust column.  However, the inferred extinction curve shows a prominent flat component between 2000-3000 \AA\ in the rest-frame, inconsistent with any locally observed template but well-fit by models of dust formed by supernovae.  Time-dependent fits to the extinction profile reveal no evidence of dust destruction and limit the decrease in the extinction column to $\Delta A_{3000} < 0.54$ mag after $t = 50$~s in the rest frame.  Our observations provide evidence of a transition in dust properties at $z \sim 5$, in agreement with studies of high-$z$ quasars, and suggest that SN-formed dust continues to dominate the opacity of typical galaxies at this redshift.
\end{abstract}


\section{Introduction}

Starting with the discovery of the 9th magnitude afterglow of gamma-ray burst (GRB) 990123 \citep{Akerlof+1999}, the early-time study of GRB afterglows has presented great promise to elucidate both the nature of the gamma-ray burst phenomenon itself and of the medium surrounding these objects in extremely distant galaxies.   Fast-responding telescopes, slewing to the burst location in time to catch the afterglow at or near the time of peak luminosity, can probe the physics of the explosion in the initial seconds as the ultrarelativistic outflow is decelerated by the interstellar medium.  Continued observations can then follow the evolution of the reverse and forward shocks for many hours as the afterglow fades away, providing constraints on the still poorly-understood early-time emission processes.  In addition, the extreme luminosities at early times \citepeg{Kann+2007,Bloom+2009,Racusin+2008} enable even very small telescopes to provide precise photometric and occasionally spectroscopic measurements of the afterglow spectral energy distribution (SED) and act as a probe of interstellar gas and dust out to the epoch of reionization \citep{Kawai+2006,Totani+2006,Gallerani+2008,McQuinn+2008,Greiner+2009a,Tanvir+2009,Salvaterra+2009}.  And while the usage of early-time SEDs as probes of the interestellar environment is hindered to some extent by the uncertain emission processes acting at these times, they nevertheless can provide constraints on the direct influence of a GRB on its surrounding medium in the form of dust destruction and photoionization \citep{Waxman+2000,Fruchter+2001,Draine+2002,Perna+2002,Perna+2003}.

At the same time, however, the fleeting and time-variable nature of GRB afterglows poses several challenges for these early-time diagnostics.   To maximise sensitivity, the smallest telescopes typically do not employ filter systems and therefore give minimal frequency-domain information.  When filters are employed, ordinary telescopes are forced to employ a filter cycle, creating the possibility of confusion between spectral and temporal evolution of the event.  Nevertheless, progress has advanced steadily with the commissioning of several simultaneous-colour robotic telescopes.  The Peters Automatic Infrared Imaging Telescope (PAIRITEL; \citealt{Bloom+2006}), online since late 2004, provides simultaneous measurements in the $J$, $H$, and $K_s$-bands every 7.8~s starting within 1--3 minutes of a typical GRB and is the primary subject of this paper.  Notable PAIRITEL-followed bursts include GRBs 041219A, 061126, and 080319B \citep{Blake+2005,Perley+2008a,Bloom+2009}.  More recently, the seven-channel Gamma-Ray Burst Optical/Near-Infrared Detector (GROND, \citealt{Greiner+2008}) has also produced simultaneous SEDs of afterglows at over the wavelength range 4000-24000 \AA, including in several cases time-dependent SEDs during the afterglow rise and fall \citep{Kruhler+2008,Kruhler+2009b} and short-timescale flares \citep{Kruhler+2009a,Greiner+2009}, and RAPTOR-T has tracked spectral changes during the fading of GRB 080319B in several optical bands simultaneously \citep{Wozniak+2009}.  In all cases, colour evolution appears to be absent or modest, consistent with the lack of strong colour evolution in the generally less constraining measurements by filter-cycling instruments such as the \Swift UVOT \citep{Oates+2009}.  Correlation with the gamma-ray prompt emission and with X-ray flares (also thought to be associated with the prompt phase: \citealt{Kocevski+2007}, \citealt{Chincarini+2007}) is rare \citep{Yost+2007}, but has been observed in some cases \citep{Vestrand+2005,Vestrand+2006,Page+2007,Kruhler+2009a,Racusin+2008,Klotz+2009}.  These multi-band observations are particularly important for distinguishing the predictions of different models for the large variety of light curve behaviors observed at early times: reverse shock \citep{Sari+1999}, energy reinjection \citep{Rees+1998}, prompt emission \citepeg{Kumar+2000,Kumar+2008}, outflow deceleration \citep{Sari+1999b,Meszaros+2006}, spectral breaks moving through the optical bandpass \citep{Sari+1998}, and many others.

GRB~071025, detected by the \Swift mission \citep{Gehrels+2004}, provides among the best probes of the early-time behavior of a gamma-ray burst to date.  While no secure spectroscopic redshift was attained ($z \sim 5.2$ was estimated from a low-quality HIRES optical spectrum at Keck; \citealt{Fynbo+2009}), the photometric SED presented here shows clear evidence of a Lyman-$\alpha$ break in the observer-frame $R$-band and indicates a photometric redshift of $4.4 < z < 5.2$ (\S \ref{sec:photoz}), making this among the highest-redshift bursts to date and one of only a few observed in simultaneous colours during prompt emission.  Our infrared and optical observations start at $\sim$30~s after the burst in the rest frame and follow the rise, peak, and fall of an afterglow in simultaneous rest-frame optical colours.  In this paper, we use this unique data set to test various models for the origin of the early emission and conclude it is likely due to the deceleration of the burst outflow into a uniform-density interstellar medium, allowing estimation of the Lorentz factor $\Gamma$ (\S\ref{sec:rise}).  The $IYJHK$ spectral energy distribution demonstrates the existence of a significant dust column obscuring a star-forming region at $z>4.4$ and provides evidence that the dust at this epoch had different properties from dust that prevails along sightlines in the more nearby universe, in agreement with the study of a high-$z$ QSO by \cite{Maiolino+2004}.  We suggest that this difference is reflective of an absence of evolved AGB stars in these earliest epochs, and search for (and place stringent limits on) signs of destruction of this dust by radiation from the GRB (\S\ref{sec:colevol}).  Throughout the paper we use the convention $F \propto t^{-\alpha}\nu^{-\beta}$ and assume cosmological parameters $h$ = 0.71, $\Omega_\Lambda$ = 0.7, $\Omega_M$ = 0.3.

\section{Observations}

\subsection{Swift}

At 04:08:54 UT on 2007 October 25\footnote{This trigger time will be used as the reference time in the remainder of the paper.}, the Burst Alert Telescope (BAT, \citealt{Barthelmy+2005}) on-board \Swift detected GRB~071025 and performed a rapid slew to the GRB location, beginning observations with the XRT \citep{Burrows+2005} at 146~s after the trigger.  The BAT light curve is broad and only slowly variable: the flux rises slowly during the first $\sim$80~s and peaks several times before beginning a steady decay at approximately $\sim$130~s.  The GRB remains detectable above the background until \Swift was forced to slew away from the position due to an Earth constraint at 422~s after the initial trigger.  Observations resumed at 3500~s, and tracked the afterglow using the XRT with no further large gaps in temporal coverage for the next $\sim$3~days, after which it became too faint to be detected.  Details of our high-energy reduction pipeline are described in detail by \cite{Butler+2007} for the \Swift BAT and by \cite{ButlerKocevski2007} for the \Swift XRT\footnote{\Swift bursts occurring after these publications, including GRB~071025, have been processed using the same methods; these results are available online at http://astro.berkeley.edu/$\sim$nat/swift/}.

\Swift's Ultra Violet-Optical Telescope (UVOT, \citealt{Roming+2005}) observed the field starting at 155 s, but detected no significant afterglow signal in any of its seven filters \citep{GCNR97.1}.  The non-detection is consistent with our photometric redshift, as outlined in \S \ref{sec:photoz}.

\subsection{PAIRITEL Observations}

The robotic infrared observatory PAIRITEL consists of the 1.3-m Peters Telescope at Mt.\ Hopkins, Arizona, formerly used for the Two Micron All Sky Survey \citep[2MASS;][]{Skrutskie+2006}, re-outfitted with the southern 2MASS camera.  PAIRITEL, like 2MASS, makes use of two dichroics to image in the $J$, $H$, and $K_s$ filters simultaneously.

PAIRITEL responded to the initial Gamma-ray Burst Coordinate Network (GCN, \citealt{Barthelmy+1995}) alert at 74.3~s and slewed immediately to the source.  Observations began at 162~s and continued uninterrupted until 3812~s, when due to a problem with the observing queue PAIRITEL temporarily slewed to another location.  Observations resumed at 9108~s and continued for another two hours.   Raw data files were processed using standard IR reduction methods via PAIRITEL Pipeline III and resampled using SWarp \citep{Bertin+2002} to create final 1.0-arcsec/pix images for final photometry.  PAIRITEL's standard observing cycle is to take three 7.8~second exposures in immediate succession at each dither position.  While the early afterglow is detected in even the shortest 7.8~second frames, the S/N was too low for reliable photometry, so the shortest exposures reported here consist of 23.4-second ``triplestacks'', the sum of all three images at each dither position.  These images were further binned at successively later times to further improve the S/N.  The afterglow position, relative to 2MASS astrometric standards, is $\alpha=355.0711583$, $\delta=+31.778575$ (J2000).

Photometry was performed in IRAF\footnote{IRAF is distributed by the National Optical Astronomy Observatory, which is operated by the Association of Universities for Research in Astronomy (AURA) under cooperative agreement with the National Science Foundation.} using the \texttt{phot} task.  Best results were achieved using aperture photometry with an aperture radius of 2.25\arcsec\ in $J$-band, 2.5\arcsec\ in $H$-band, and 2.75\arcsec\ in $K_s$-band.   Unfortunately, while conditions during the observations were generally clear, the night was not fully photometric, with variations in the transmission of up to 0.3 mag during the course of observations and additional significant fluctuations in the seeing.  Calibration was therefore performed by re-determining the zeropoint for each image individually by comparison to our secondary field standards (\S \ref{sec:calib}).  Fortunately, the field of GRB~071025 is rich in bright field stars, and a total of eight nearby stars (present and well-detected in even short exposures with reference uncertainties of $<$0.05 mag) were used to determine the zero-point.  The statistical uncertainty on the zero-point (never more than 0.05 mag) is essentially negligible relative to other sources of error in all cases.   Systematic sources of error are addressed in \S \ref{sec:ircal}.

The large plate scale of PAIRITEL (2.0 \arcsec/pix) and the variable sub-pixel response function of the NICMOS3 arrays creates a significant additional uncertainty in each position beyond ordinary photometric errors, estimated at $\sim$3 percent by \cite{Blake+2008}.  To quantify this uncertainty as accurately as possible, we constructed light curves for standard stars of different magnitudes in regions of the image free of defects by measuring the image-to-image magnitude variations of bright (source-dominated) stars.  An additional uncertainty of approximately $\sim$0.02 mag per position was required to incorporate the observed scatter in the photometry of these objects.  Additionally, we examined fainter (sky noise-dominated) stars to compare the IRAF-generated uncertainty to that observed in the zero-pointed light curve, finding the IRAF uncertainties to be too low by about 20\% in each filter.  Therefore the final uncertainties on our photometry, reported in Table \ref{tab:ptelphot}, were determined by increasing the IRAF uncertainty by 20\% and adding the result in quadrature with 0.02/$\sqrt{N_{\rm pos}}$ mag, where $N_{\rm pos}$ is the number of unique dither positions per stacked image. 

\subsection{REM Observations}

GRB~071025 also triggered REM (Rapid-Eye Mount; \citealt{Zerbi+2001}), a robotic \citep{Covino+2004} telescope located at the ESO Cerro La Silla observatory (Chile). The REM telescope has a Ritchey-Chretien configuration with a 60 cm f /2.2 primary and an overall f /8 focal ratio in a fast moving alt-azimuth mount providing two stable Nasmyth focal stations. At one of the two foci, the telescope simultaneously feeds, by means of a dichroic, two cameras: REMIR \citep{Conconi+2004} for the NIR  and ROSS \citep{Tosti+2004} for the optical. Both cameras have a field of view of 10x10 arcmin and imaging capabilities with NIR ($1\mu$, $J$, $H,$ and $K$) and Johnson-Cousins $VRI$ filters.  Observations of the GRB~071025 field began at 144~s after the trigger, although this initial $H$-band exposure did not detect the afterglow.    The optical camera was unfortunately not operational due to maintenance, so exposures were acquired only in $1\mu$, $J$, $H$, and $K$.

The raw frames were corrected for dark, bias, and flat field following standard procedures.  Although the burst was at low elevation at the trigger time, seeing conditions were good and photometry was performed using a 3.5 pixel (1.2\arcsec) aperture.  Conditions were not photometric, and so the zeropoint was determined for each image individually in $JHK$ bands using a subset of 2MASS-based standards.  The $1\mu$-band (often referred to as $z$ in previous work, though this filter has almost no overlap with the traditional SDSS $z$-band), after taking into account the transmission of the ROSS/REMIR dichroic, is close to the MKO $Y$-band\footnote{http://www.ukidss.org/technical/instrument/filters.html; see also \cite{Hillenbrand+2002}} and so we treated this filter as a $Y$ measurement, using the interpolated magnitudes in Table \ref{tab:optcal} (see \S \ref{sec:calib}) and basing the calibration on four reference stars well-detected in all images.

\subsection{RAPTOR Observations}
\label{sec:raptor}

The RAPTOR (Rapid Telescopes for Optical Response) experiment \citep{Vestrand+2002}, operated by Los Alamos National Laboratory, consists of a series of small telescopes at the Fenton Hill Observatory in New Mexico.  RAPTOR-S is a fully autonomous robotic telescope with a 0.4-m aperture and typical operating focal ratio f/5.  It is equipped with a 1000$\times$1000 pixel CCD camera employing a back-illuminated Marconi CCD47-10 chip with 13$\mu$ pixels. 

RAPTOR-S responded automatically to the localization alert and was on target at 04:10:14.95 UT, 81.3~s after the trigger time (4.2~s after receiving the GRB position). The rapid response sequence of RAPTOR-S consists of nine 5-second images followed by twenty 10-second images and finally 170 30-second images for a total of $\sim$2 hours of coverage (including 5-second intervals between exposures used primarily for readout). In order to improve the S/N ratio, photometry was performed on coadded images. Aperture photometry was performed using the SExtractor package \citep{Bertin+2002}, and the magnitude offsets between epochs were derived using several dozen field stars.

Because of the extreme redness of this afterglow, the unfiltered RAPTOR-S observations required a special calibration procedure.  Although the effective wavelength of the response curve for RAPTOR-S is close to that of the standard $R$ band, the spectral energy distribution of this burst (\S \ref{sec:photoz}) indicates a sharp drop in the flux between $I$ and $R$ bands, likely due to the onset of the Lyman-$\alpha$ forest.  As a result, most photons detected by RAPTOR-S actually fall in the spectral region covered by the standard $I$ filter.

Therefore, we tie the unfiltered data to $I$-band standards from the Lick calibration (\S \ref{sec:calib}).  The offset $(m_{\rm C}-I)_{\rm star}$ between the unfiltered magnitudes and standard $I$ was derived using 7 well-measured stars in the vicinity of the GRB covering a narrow range of colours $0.5 < (R-I) < 0.66$.  Assuming that the SED of the burst emission did not change significantly between the time of RAPTOR-S observations and the time when it was measured, we derived an approximate correction to $(m_{\rm C}-I)_{\rm star}$ to account for the extremely red colour of the GRB. We used a K5V model spectrum from \cite{Kurucz1979} as a proxy SED matching the mean colour of our comparison stars. By folding both SEDs with response curves of RAPTOR-S and the standard $I$-band filter we find $(m_{\rm C}-I)_{\rm GRB} = (m_{\rm C}-I)_{\rm star} + 0.74$ mag.  The uncertainty of the derived zero point is about 10\%; consistent with this, we measure a relatively small offset of $-0.08$ magnitudes between the calibrated RAPTOR magnitudes and an extrapolation from later, filtered $I$-band observations using our light curve model (see \S \ref{sec:lc}).   Table \ref{tab:phot} lists the final RAPTOR-S photometry.

\subsection{Super-LOTIS Observations}

Super-LOTIS (Livermore Optical Transient Imaging System) is a robotic 0.6-m telescope dedicated to the search for optical counterparts of gamma-ray bursts \citep{Williams+2004,Williams+2008}. The telescope is housed in a roll-off-roof facility at the Steward Observatory Kitt Peak site near Tucson, Arizona.  Super-LOTIS triggered on GRB~071025 and began observations at 04:10:29 UT (95~s after the trigger), acquiring a series of $R$-band frames, which were reduced using standard methods.  Unfortunately, because of the optical faintness of the afterglow and high sky background, the quality of the images is poor and even after extensive stacking the detection is marginal, particularly in the earliest few stacks.  Photometry was performed using aperture photometry and our Lick $R$-band field calibration as detailed in \S\ref{sec:calib}.

\subsection{Lick Infrared Observations}
\label{sec:lickir}

We acquired an additional series of infrared observations using the 3m Shane telescope at Lick Observatory equipped with the UCLA GEMINI IR camera \citep{McLean+1993,McLean+1994}.  A total of nine exposures were acquired in $J$ and $K'$ bands simultaneously starting at 04:52:23 UT, integrating for 11 coadds of 20~s each in $J$ and 35 coadds of 6~s each in $K'$.  The IR afterglow was still very bright at this time, and is clearly detected with signal-to-noise $S/N > 50$ in individual exposures.  Reduction was performed via direct subtraction of temporally adjacent exposures followed by division by a twilight flat.  Photometry was performed using IRAF and an aperture of 2 pixels ($1.4 \arcsec$); images were calibrated relative to the PAIRITEL magnitudes of five nearby bright field stars.

The response of the $K'$ filter is significantly different from $K_s$ and the GRB exhibits an apparent colour ($H-K \approx 1.0$ that is much redder than any field star used for comparisons (ranging between $H-K = 0.04-0.18$).  To correct to $K_s$ for direct comparison to the PAIRITEL data, we use an approximate correction of $K_s \approx K \approx K' - 0.07$ \citep{Wainscoat+1992}, with this correction inferred from the reddest star in Table 1 of that work (Oph S1, $H-K = 0.94$, $K'-K = 0.07^{+0.015}_{-0.025}$).  The $K$ to $K_s$ colour term is assumed to be negligible.  This is found to produce good agreement between Lick data and coeval PAIRITEL points.  However, due to uncertain differences between the Lick, MKO, and other filter sets and the intrinsic GRB spectrum itself the overall calibration offset could be as much as $0.05$ mag, and as a result the Lick $K$ photometry is not used in fitting.

\subsection{MAGNUM Observations}

The MAGNUM (Multicolour Active Galactic NUclei Monitoring) 2.0~m telescope on Haleakala has been carrying out observations of AGN and other variable objects (including GRBs) since 2001 \citep{Yoshii+2002,Yoshii+2003,Kobayashi+2003}.  The telescope is equipped with dual optical and infrared channels, allowing simultaneous observations in two bands.

We initiated MAGNUM observations starting at 06:59 UT, acquiring a sequence of dithered exposures over the next $\sim$2 hours in a large number of filters, including $RI$ in the optical channel and $YJHK$ in the infrared channel.  The MAGNUM FOV is small, and generally only one star was present in the field and away from the chip edge at all dither positions.  Therefore only a single star was used to establish the calibration in each filter.  In the $H$, $K$, and $Y$-band observations the star at $\alpha$=355.066002, $\delta$=31.793428 was used for this purpose; for $R$, $I$, and $J$ the star at $\alpha$=355.058815 $\delta$=31.780569 was used.  The second $Y$-band exposure unfortunately contained no usable reference star.  However, comparison of exposures in other filters and at other points in the night suggest that conditions were photometric, and so calibration was achieved by comparison to the first Y-band exposure (with a small aperture correction.)

\subsection{Kuiper Observations}
\label{sec:kuiper}

Shortly after the GRB trigger, we initiated imaging observations at the 1.54m Kuiper telescope, operated by Steward Observatory and located on Mt. Bigelow.  Observations began at 04:37:08 UT and continued until 08:55:41 UT, mostly in the $R$ and $I$ filters with some additional observations in $V$.  Images were reduced and combined in IRAF using standard techniques.  The $I$-band images were not dithered and so an archival fringe frame was used to subtract the fringe pattern.  Photometry was performed in IRAF using secondary standards.

\subsection{Late-Time Afterglow Observations}

To try to constrain the late-time ($t > 12$ hr) behavior of this burst, additional follow-up was carried out on the 3.6-m New Technology Telescope and at GROND.  We observed the burst location on NTT using the infrared imager SOFI in a series of $J$, $H$, and $K$-band exposures, and additionally in $H$-band only on the following night.  Photometry was calibrated relative to our IR secondary standards.

GROND is a seven-channel instrument which has been mounted on the ESO 2.2-m telescope at La Silla, Chile, since April 2007.  GROND began observations of GRB~071025 on 2007 Oct 26 at 01:50 UT and completed one 8 minute observing block and two 20 minute observing blocks.  In total, 9 images were taken in the $g'r'i'z'$ bands and 216 were taken in the NIR. Each NIR image was 10~s long; the optical images varied in length from 137 to 408~s.  The images were reduced using the GROND pipeline \citep{Yoldas+2008}, with all images combined into a single stack for each filter.   For consistency with other measurements, photometry was performed using aperture photometry calibrated to our secondary standards in $JHK$.  For $g'r'i'z'$ bands, images are calibrated directly relative to spectroscopic standard stars SA 114-750 and SA 114-656.

Poor agreement is observed between the NTT and GROND observations (and between the overall SED at this time and earlier data) using a standard 1\arcsec aperture, even though these epochs are effectively coeval.   We have re-examined these data and find no clear evidence of problems in the reduction or photometry, although the afterglow appears extended in the N-S direction in the GROND $H$-band frame, suggesting that it might be blended with a nearby source or image artifact.  No neighboring source is observed in the NTT imaging, and the deep Keck optical imaging shows no object within $\sim$3$\arcsec$ of the afterglow position (\S\ref{sec:keck}).  However, to guard against this possibility we performed the photometry in the GROND $J$ and $H$ channels and all NTT channels using a small aperture ($0.5$\arcsec) in all bands.   This smaller aperture provides good consistency between the two observations and is used in our analysis.

\subsection{Keck Observations}
\label{sec:keck}

To help rule out a low-redshift origin for this burst we imaged the field around GRB~071025 with LRIS \citep{Oke+1995} on the Keck I telescope on 2008 Aug 02 using the $g$ and $R$ filters simultaneously under excellent conditions.  Total exposure times were 1050~s in $R$-band and 1140~s in $g$-band.  Consistent with the large photometric redshift inferred from the SED, no significant flux due to a host galaxy was detected at the location of the optical/IR afterglow (the nearest object is a pair of faint point-like sources located 3\arcsec\ to the northeast).  Forced photometry at the position of the optical afterglow, calibrated using unsaturated secondary standards, gives a limit (3-$\sigma$) of $R > 26.5$ mag, $g > 27.0$ mag.

\subsection{Field Calibrations}
\label{sec:calib}

To improve upon the photometric accuracy of 2MASS, we stacked together all observations of the GRB field acquired by PAIRITEL during the night of 2007 October 25 UT and calibrated a set of isolated, high-S/N stars present in the field in all or nearly all dither positions relative to 2MASS.  These magnitudes were used in place of 2MASS magnitudes directly and are presented in Table \ref{tab:ircal}.

For the optical filters, on the night of 2009 June 19 we observed the field of GRB~071025 using the Nickel 1m telescope at Lick observatory.  Conditions were photometric throughout the night.  Three exposures were acquired in $R$-band and one each in $I$ and $g$ band and stars within the field were calibrated by comparison to repeated observations of PG 1633 and PG 2336 \citep{Landolt1992} at varying airmass, calibrating reference stars within the field.  A second calibration was conducted on 2009 Sept 28 using repeated observations of standard fields PG 1633, PG 2336, PG 0213, and SA 110; the results were found to be completely consistent with the June calibration.

No field calibration was performed in the $Y$-band.  To calibrate the observations in this filter, we derived our own transformation equation for calculating $Y$ magnitudes of reference stars given photometry in nearby bands by fitting a simple linear regression model to the photometry available online at the UKIRT webpage\footnote{http://www.jach.hawaii.edu/UKIRT/astronomy/calib/phot\_cal/fs\_izyjhklm.dat}.  (The $Y-J$ colour was fit as a linear function of $J-H$, and the residuals were then fit to a linear function of $I-J$.)  The transformation equation $Y = J + 1.104(J-H) - 0.11(I-J) - 0.03$ was found to accurately describe the observed $Y$-band magnitudes for the available standards (with photometry in all four bands) with an RMS of $<0.03$ mag.  We therefore applied this equation to calculate the $Y$ magnitudes for secondary standards in the GRB~071025 field using the calibrated $IJH$ photometry.

The final calibrated magnitudes for these stars are presented in Table \ref{tab:optcal}.

\section{Analysis}

\subsection{Early-Time Afterglow Evolution: Rise, Fall, and Reddening}
\label{sec:lc}

\begin{figure}
\centerline{
\includegraphics[scale=0.75,angle=0]{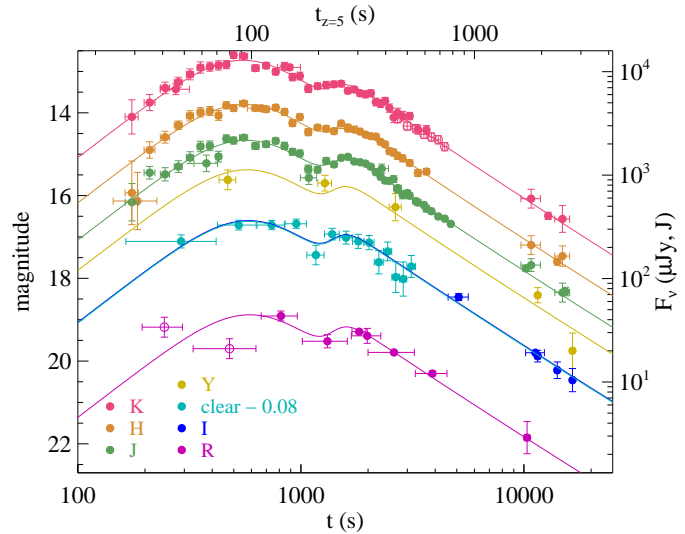}}
\caption{Early-time multiband optical and infrared light curves of GRB 071025 fit to our empirical light curve model.  The afterglow is caught during its rise at $\sim$30~s in its rest frame (assuming $z  = 5$), and exhibits a double-peaked structure before fading again as a simple power-law.  The RAPTOR unfiltered data has been shifted to match the I-band data.  Magnitudes are Vega-based and not corrected for extinction.}
\label{fig:lcurve}
\end{figure}

All photometric observations of GRB~071025 during the first night are presented in Figure \ref{fig:lcurve}.  Several features are immediately apparent.  First, the afterglow was caught during what appears to be its initial optical rise, brightening by $\sim$1.5 mag from the first detections to the peak in all filters.  Second, the evolution is not single-peaked: a limited rebrightening is observed at $\sim$1800~s.  Third, the burst is extremely red, with $R-K \sim 6.5$ mag.   Finally, no dramatic colour change is evident.  This is not to say that there is not finer-scale colour evolution, however---as will be discussed later, the best-fit curves plotted in Figure \ref{fig:lcurve} correspond to a chromatic model which is shown to produce a large improvement in $\chi^2$ relative to the monochromatic case.

\begin{table}
\begin{minipage}{80mm}
\centering
\caption{Light Curve Best-Fit Parameters}
\begin{tabular}{lll}
\hline
Parameter     &  symbol           & value \\
\hline
C1 rising index             &$\alpha_{\rm 1,r}$       &$-1.66 \pm 0.15 $  \\
C1 fading index             &$\alpha_{\rm 1,f}$       &$ 1.73 \pm 0.21 $  \\
C2 rising index             &$\alpha_{\rm 2,r}$       &$-11.0 \pm 2.1  $  \\
C2 fading index             &$\alpha_{\rm 2,f}$       &$ 1.27 \pm 0.04 $  \\
C1 peak time (s)            &$t_{\rm pk,1}$           &$ 575  \pm 42 $    \\
C2 peak time (s)            &$t_{\rm pk,2}$           &$ 1437 \pm 17 $    \\
Ratio of C2/C1 peak flux    &$F_2$                    &$ 0.24 \pm 0.03 $  \\
Colour change across C1 peak&$\Delta\beta_{\rm 1(rf)}$&$-0.20 \pm 0.14 $  \\
Colour change between C1/C2 &$\Delta\beta_{\rm 12}$   &$-0.26 \pm 0.12 $  \\
\hline
Flux at $t=10000$s          &$F_R$                    &$5.76  \pm 1.15 $  \\
                            &$F_I$                    &$34.4  \pm 3.48 $  \\
                            &$F_Y$                    &$84.45 \pm 8.66 $  \\
                            &$F_J$                    &$118.9 \pm 4.79 $  \\
                            &$F_H$                    &$155.4 \pm 6.26 $  \\
                            &$F_K$                    &$250.4 \pm 15.1 $  \\
\hline
\end{tabular}
{Summary of free parameters fit in the light curve model.  Peak times are for the $J$-band filter.  Flux parameters are not corrected for Galactic extinction; uncertainties include added systematics. ``C1'' refers to the first light-curve component; ``C2'' refers to the second component.}
\label{tab:lcpar}
\end{minipage}
\end{table}

\begin{figure}
\centerline{
\includegraphics[scale=0.75,angle=0]{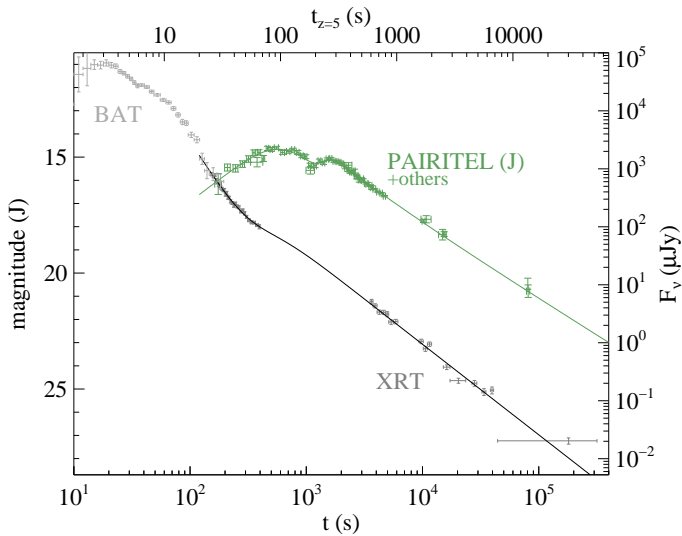}}
\caption{The gamma-ray (\Swift BAT) and X-ray light curves (\Swift XRT) of GRB 071025, compared to the $J$-band light curve out to late times.  The X-ray light curve is rapidly fading during the optical-IR rise, probably due to high-latitude prompt emission (the light curve connects smoothly with the BAT light curve at these times if scaled to the X-ray flux, as shown.)  Both optical and X-ray light curves fade with an unbroken decay at late times but with different decay slopes: $\alpha_{\rm opt} = 1.27 \pm 0.04$ versus $\alpha_{\rm X}= 1.56 \pm 0.03$.}
\label{fig:jxlc}
\end{figure}

The empirical model used to fit this burst is described in previous work \citep{Perley+2008b,Bloom+2009}.  In brief, our method fits all data in all filters simultaneously using a series of summed \cite{Beuermann+1999} broken power-law functions, in this case two per filter.  The sharpness parameter $s$ was fixed at 1 (allowing it to vary resulted in insignificant improvement to $\chi^2$).  For each component, the pre-break (rising) power-law index $\alpha_{\rm r}$ is constrained to be the same for all filters, as is the post-break (fading) index $\alpha_{\rm f}$.  The overall flux-normalization factor in each filter is arbitrary, determined by the best fit to the data.  Colour change is modeled by assuming that any overall change takes the form of a variation in the intrinsic spectral index $\beta$, i.e., $F_{\nu,2} = F_{\nu,1} \times (\frac{\nu_2}{\nu_1})^{\Delta\beta12}$.  Colour is allowed to vary between components (simply using the above equation to tie the normalised flux parameters of each component) and between rising and falling segments of an individual component (by allowing the break time of the Beuermann function to vary as a function of frequency, using the central frequency of each filter).  It should be emphasised that this method makes no assumptions about the overall SED, since only the \emph{variation} in $\beta$ is constrained.  Indeed, the fitting method can be used to generate a best-fit observed SED in all available filters, using all available data, at any chosen time (\S\ref{sec:photoz}). 

Two components (best-fit parameters are summarized in Table \ref{tab:lcpar}) are found to provide an excellent fit to the data.\footnote{The first two Super-LOTIS points are an exception, both of which deviate from the fitted model by 2--3 $\sigma$.  Given the low signal-to-noise detections and large degree of time-binning in both cases, these points are not included in the fit, although the low flux observed in the second observational window, which covered the peak of the light curve, is nevertheless surprising given the behavior in all three PAIRITEL bands and in RAPTOR data at that time.}  The light curve brightens quickly between our first detections at 180~s with a power-law of approximately $\alpha_{\rm 1,r} = -1.66 \pm 0.15$ to a smooth peak at 580~s, then fades until about 1200~s.  At that point the afterglow briefly rebrightens, peaking again at $\sim$1400~s before fading as a simple power-law ($\alpha_{\rm 2,f} = 1.27 \pm 0.04$) for the remainder of our observations.  The $\chi^2$ residual, assuming no colour change, is 222.4 per 154 degrees of freedom (dof).  Permitting colour change improves the fit significantly:  allowing the parameters $\Delta\beta_{\rm 12}$ (describing the change in intrinsic spectral index between the first and second component after peak) and $\Delta\beta_{\rm 1(rf)}$ (describing the change in intrinsic index between rising and falling portions of the first component) to both vary, $\chi^2$/dof improves to 197.4/152 which (according to the $f$-test) is significant at $>99.9$\% confidence.   Most of this change is associated with the transition to the second component ($\Delta\beta_{\rm 12} = -0.26 \pm 0.12$, versus a not-significant colour change across the first peak of $\Delta\beta_{\rm 1(rf)} = -0.20 \pm 0.14$) but unfortunately, although the need for overall red-to-blue colour change is clear, its nature cannot be clearly distinguished by this methodology.  We will further examine scenarios for this possible colour change in \S\ref{sec:colevol}.

The X-ray light curve (Figure \ref{fig:jxlc}) was fit using a similar method (but with only a single ``filter'', simplifying the process significantly).  Again, two summed functions are found to provide an acceptable fit to the data.  However, the first component is a rapidly-declining, unbroken power-law with $\alpha_{\rm X,init}=3.1\pm0.2$.  This initial segment connects smoothly with the BAT prompt emission, as has been seen for a large majority of \Swift bursts \citep{OBrien+2006}.  The optical peaks unfortunately fall during an orbital gap in the XRT coverage, but by the end of the first observations the power-law is already clearly flattening, almost certainly due to the transition from the rapid decay phase \citep{OBrien+2006} to a standard afterglow \citep{Nousek+2006}.  Coverage resumes approximately an hour later, by which time the X-ray light curve is fading rapidly in an unbroken decay with $\alpha_{\rm X}=1.56 \pm 0.03$.

\subsection{SED and Photometric Redshift}
\label{sec:photoz}

At 10000~s after the burst, the evolution of the light curve has given way to a simple power-law decay dominated by only a single component.  Moreover, thanks to the MAGNUM observations, photometry is available in all colours within a relatively short time span surrounding this epoch with high S/N in $JHK$.  We therefore choose this time as the extraction point for the overall spectral energy distribution (SED) of this burst, using the model fluxes from our fit as described above. (These fluxes are consistent with the MAGNUM and PAIRITEL photometry measured at this epoch specifically.)  All fluxes are corrected for Galactic extinction (relatively small at $E_{B-V} = 0.07$ mag in this direction; \citealt{Schlegel+1998}.)

\begin{figure}
\centerline{
\includegraphics[scale=0.75,angle=0]{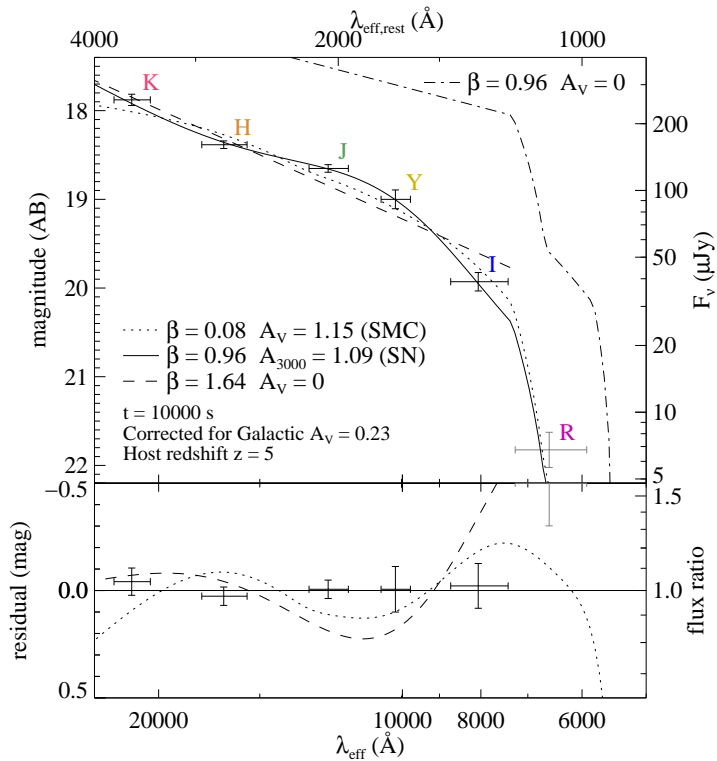}}
\caption{Spectral energy distribution of GRB 071025 inferred from our broadband photometry, fit with different extinction models.  Note the spectral flattening between $J$ and $H$ that contrasts with red $H-K$ and $I-Y$ colours.  (The $R-I$ colour is due to absorption by the Lyman-$\alpha$ forest.)  Traditional models (such as SMC-like extinction, shown here as a dotted line) cannot reproduce this feature and give poor fit residuals ($\chi^2$/dof$ = 20.8 / 2$).  The supernova-dust model of Maiolino et al. (2004), shown as the solid line, is an excellent fit ($\chi^2$/dof$ = 0.81 / 2$).  The dot-dashed line represents the intrinsic afterglow SED (for the SN model) without extinction applied but including our model of the IGM opacity at this redshift.}
\label{fig:sed}
\end{figure}

The $1\sigma$ uncertainty on the fit parameter was combined in quadrature with an estimate of the calibration uncertainty in each filter.   In the $J$ and $H$ filters, where the afterglow is comparable in colour to reference stars (which show negligible scatter), we use an uncertainty of 0.04 mag; in $K$ where the afterglow colour is much redder than our reference stars we conservatively increase this to 0.06 mag.  This incorporates both the absolute and relative calibration accuracy of 2MASS (estimated at $\sim$0.02 and $0.011$ mag, respectively; \citealt{Cohen+2003} and 2MASS online documentation\footnote{http://www.ipac.caltech.edu/2mass/releases/allsky/doc/sec4\_8.html}), effects of variation of the effective wavelength $\lambda_{\rm eff}$ from its reference value due to a non-standard spectrum ($< 0.02$ mag), the possibility of strong absorption from ISM or IGM lines (very likely $< 0.02$ mag), and uncertainties in the extinction correction ($< 0.01$ mag).  In $Y$-band, we use an estimate of the photometric scatter of the high-S/N REM reference stars to the interpolated secondary standards (0.1 mag).  We also use 0.1 mag in $I$-band due to the redness of the afterglow in this band and the possibility that Lyman-$\alpha$ may be affecting the flux towards the blue filter edge if the redshift is $z>5.0$.  In $R$-band a large uncertainty of 0.2 mag is used, although because $R$ is almost certainly heavily blanketed by the Ly-$\alpha$ forest we use this filter only to place a limiting value on the redshift and exclude it from fits to the extinction profile.  The resulting SED (fit with various models, explained below) is plotted in Figure \ref{fig:sed}.

The sharp dropoff towards the $R$-band is suggestive of high redshift.  However, the spectral slope observed even well redward of this apparent break is quite red ($\beta \sim 1.64$, as shown by the dashed straight line in Figure \ref{fig:sed}), suggesting that significant extinction is likely present as well.  In order to quantitatively constrain the redshift $z$, we fit the data set with a large number of different extinction models (detailed in \S \ref{sec:ext}) at varying redshifts.  Absorption due to the Lyman-$\alpha$ forest is taken into account using a simple model of the average opacity of the IGM as a function of $z$ and $\lambda$ from \cite{Madau+1995}.  The extinction column $A_V$ and the spectral index $\beta$ were constrained to be positive: negative extinction is unphysical, while a negative spectral index would be both much bluer than any previously observed afterglow and in disagreement with standard afterglow theory \citep{Sari+1998}.

The HIRES spectrum discussed in \cite{Fynbo+2009} shows a trace extending from the limit of the spectral range at 7950 \AA\ down to 7550 \AA, blueward of which no flux is detected.  While the quality of this spectrum is poor, the nondetection of Lyman-$\alpha$ puts a robust upper limit on the redshift of $z<5.2$, so this was treated as the maximum redshift.  Regardless of the extinction law, no known dust curve is able to reproduce the extremely steep $I-R$ slope without invoking Lyman-$\alpha$ blanketing of the $R$-band, which becomes significant at $z \sim 4.0$.   Even after including a variety of extinction templates (below), the lower limit on the redshift (95\% confidence) is $z>4.4$.  Treating redshift as a free parameter, the best-fit $z$ is dependent on the extinction law but is approximately $z=4.8 \pm 0.2$ ($1\sigma$).  In the remaining discussion we will assume a fiducial value of $z=5.0$; however, similar conclusions apply to other redshifts within the constrained range ($4.4-5.2$).

\subsection{Extinction Profile}
\label{sec:ext}

Qualitatively, the SED presented in Figure \ref{fig:sed} is unusual among GRB afterglows due to the presence of an apparent inflection:  while the $K-H$ and $Y-I$ colours are very red, between $H$ and $J$ bands the slope is quite flat.  This flattening is quite significant (e.g., the $H$-band point is more than 0.2 mag below an interpolation between $J$ and $K$) and suggests that the afterglow of GRB~071025 is subject to a complex reddening profile.   To try to distinguish different possible models, we therefore fit many different extinction laws to the photometric SED, including Milky Way, LMC, and SMC curves estimated using the parametrization of \cite{Fitzpatrick1999} as implemented in the GSFC IDL astronomy user's library, the starburst-galaxy Calzetti curve \citep{Calzetti+2000}, and the high-$z$ QSO extinction law from \cite{Maiolino+2004}.  The intrinsic spectral index $\beta$ is free but limited to be $\beta > 0$.  A summary of the goodness-of-fit $\chi^2$ for each fit model is presented in Table \ref{tab:extfits}.

A large family of models, including the Milky Way and Large Magellenic Cloud curves as well as the extinction curves derived from a few recent highly reddened GRBs \citep{Kruhler+2008, Prochaska+2009, Eliasdottir+2009}, display a prominent 2175~\AA\ bump.  We can strongly rule out such a feature:  at the observed redshift, the broad absorption signature would fall in or near the $J$-band.  Formal fits using these extinction templates (regardless of $R_V$) return $A_V$ of zero in all cases (our fits do not permit negative extinction.)

An SMC-like extinction curve (dotted line in Figure \ref{fig:sed}) provides a visually reasonable-looking fit to our data, but the $\chi^2/$dof is unacceptable at 20.8/2.  This is again no surprise:  the SMC extinction curve increases rapidly and monotonically with decreasing wavelength and cannot produce the flattening in our SED.  The featureless Calzetti law similarly produces a poor fit because it cannot produce the deviations from a power-law evident in the photometry.  

We also attempted a general fit using the full parameterization of \cite{Fitzpatrick1999}, but even if the $\gamma$ and $x_0$ parameters of this model are fixed and the $c_1$ and $R_V$ parameters are tied $c_2$ using e.g. the correlations of \cite{Reichart2001}, the solution is underdetermined.  If the intrinsic spectral slope $\beta$ is fixed, the solution is exactly determined; for e.g. $\beta=0.65$, we derive $R_V=5.26\pm0.53$, $c_2=0.17\pm0.12$, $c_3<0.2$, $c_4=1.03\pm0.32$, $\chi^2/$dof = 1.49/0.  However, this combination of parameters (small $c_2$ and low or zero $c_3$, indicating a shallow near-UV extinction law and negligible 2175\AA\ bump) is unlike any sightline in the local universe observed to date.  We also fit the data to the general extinction curve of \cite{LiA+2008}, fixing $\beta=0.65$ and $c_4=0$ to avoid underdetermination, but the $c_1$ and $c_3$ parameters did not converge.

However, one previously observed extinction law performs extremely well at matching the observed features.  \cite{Maiolino+2004} presented observations of the reddened $z=6.2$ broad absorption line quasar SDSSJ104845.05+463713, comparing NIR spectroscopy of the source to optical spectra of low-redshift quasars of the same class to estimate the extinction law.  The inferred curve of this object is notable for a distinct flattening between $1800-3000$\AA, and was interpreted (and modeled quantitatively) by that paper as the signature of dust synthesised in supernova explosions.  We fit a polynomial to the solid ($Z = 10^{-4} Z_{\odot}, M = 25 M_{\odot}$) curve displayed in Figure 2 of that paper and used the resulting extinction curve to fit our observed photometry.\footnote{Our $K$-band point is not covered by this figure, as the corresponding rest wavelength is shifted out of the IR window at $z=6.2$.  We assume an approximately linear extinction law in $1/\lambda$ below $\lambda_{\rm rest} < 3300$ \AA.}  The result is an excellent match ($\chi^2/$dof = 0.81/2) and is shown as the solid line in Figure \ref{fig:sed}.  The associated extinction column is $A_{3000} = 1.09 \pm 0.20$ mag.\footnote{The Maiolino extinction curve is normalised to $A_{3000}$ instead of $A_{V}$ (the $V$-band at $z>5$ is shifted into the mid-IR).}  

The best-fit value of the intrinsic spectral index $\beta_{\rm IR}$ as inferred at the SED extraction epoch is $\beta_{\rm IR}=0.94 \pm 0.14$, quite typical of other afterglows at this stage.  This value is also consistent (albeit only marginally) with the theoretically expected value based on the observed X-ray spectral index (intrinsic $\beta_{\rm X} = 1.15 \pm 0.12$) if a cooling break is present between IR and X-ray bands: in this case  $\beta_{\rm IR}$ = $\beta_{\rm X} - 0.5$ = $0.65 \pm 0.12$.)  Imposing this constraint as a prior on the fit to $A_V$ and $\beta_{\rm IR}$, we measure $A_V = 1.27 \pm 0.20$ mag.

Alternatively, the SED is also consistent with the presence of no cooling break: at the extraction epoch the combined IR-through-X-ray SED is well-fit ($\chi^2$/dof = 1.14/3) by a single power-law with $\beta_{\rm IR,X} = 0.88$ and $A_V = 1.19 \pm 0.20$ of Maiolino dust, both consistent with the values inferred from the optical data alone.  However, the X-ray flux at this time is clearly fading faster than the optical light curve (Figure \ref{fig:lcurve}): if this is not due to the presence of a moving spectral break such as a cooling break, the spectral index itself would have to be slowly evolving (implying evolution in the electron index $p$).

In support of our general conclusion of a significant amout of dust extinction, we note that a large amount of absorption is inferred from the X-ray spectrum also: we measure an equivalent column of $N_{\rm H} = (3.2 \pm 0.8) \times 10^{22}$ cm$^{-2}$.  Although the scatter in the ratio of $A_V$/$N_{\rm H}$ for \Swift bursts is nearly an order of magnitude, using the average value from \cite{Schady+2007} this column corresponds to an extinction of $A_V \sim 4$ mag.

\begin{table}
\begin{minipage}{80mm}
\centering
\caption{Results of Extinction Fits}
\begin{tabular}{lllll}
\hline
Dust Model &  $\beta$ & $A_V$ &$R_V$ & $\chi^2$ $/$ dof \\
           &          & mag   &      &                  \\
\hline
none          &   $1.64 \pm 0.08$ & 0                &                  & 33.3 / 3 \\
SMC           &   $0.08 \pm 0.42$ & $0.12 \pm 20.6$  & 2.73             & 20.8 / 2 \\
MW            &   $1.64 \pm 0.08$ & $< 0.07$         & 3.1              & 33.3 / 2 \\
LMC           &   $1.64 \pm 0.08$ & $< 0.07$         & 3.2              & 33.3 / 2 \\
GRB080607     &   $1.64 \pm 0.08$ & $< 0.12$         & 4.0              & 33.3 / 2 \\
Calzetti      &   $0.00 \pm 0.80$ & $1.42 \pm 0.68$  & 4.0              & 25.3 / 2 \\
Fitzpatrick   &   $0.65$          & $2.52 \pm 0.97$  & $5.26 \pm 0.53$  & 1.49 / 0 \\
Maiolino SN   &   $0.96 \pm 0.14$ & $1.09 \pm 0.20$\footnote{Value is $A_{3000}$.}  &   & 0.81 / 2 \\
\hline
\end{tabular}
Summary of key parameters from fits of various dust models to the SED of GRB~071025 as modeled at $t$=10000~s.  A redshift of $z=5$ is assumed in all cases.
\label{tab:extfits}
\end{minipage}
\end{table}

\subsection{Further Investigations of the IR Calibration}
\label{sec:ircal}

The inferrence of SN-type dust for this object depends sensitively on the accuracy of our photometric calibration, and statements of its significance relative to the SMC fit depend equally critically on the precision in the $JHK$ bands being as good as we claim: the Maiolino model is no longer preferred at $>95$\% confidence if, for example, additional uncertainty of $>$0.075 mag (in addition to the systematic uncertainties already applied; \S\ref{sec:photoz}) is added in quadrature to all SED data points, or if $>$0.1 mag is added to just the $H$-band point (dependence on the other data points is much more robust: an addition of $>$0.2 mag to $K$ is required, and any one of the $J$, $Y$, or $I$ points could be removed completely).  Therefore we have scrutinised in detail our infrared calibration procedures with particular emphasis on the PAIRITEL data.  Because of the large number of exposures and large number of calibration stars detected at high S/N, the statistical errors on the zeropoint are small.  Possible sources of systematic uncertainty (beyond the minor effects we have already discussed and included) we have considered include:

\textbf{Instrumental colour terms.}  PAIRITEL uses the same telescope, filter, and camera system as the 2MASS survey, and so there is no reason to expect colour terms associated with the optics to be present.  However, the presence of a significant bandpass difference could cause systematic discrepancies in calibration relative to field stars (see also \S\ref{sec:lickir}), in particular in $H$ and $K$ bands where the afterglow colour is much redder than any of the bright stars used for calibration.  We inspected the magnitudes derived from stars in our deep stack as compared to the stars in 2MASS to search for a correlation between the magnitude offset and colour; none was found.

\textbf{Strong atmospheric variations in the effective filter bandpass.}    The infrared absorption bands associated with water in Earth's atmosphere exhibit time-variability, even within the observational windows.  In particular, the exact shape of the $J$-band transmission function depends on the amount of precipitable water vapor (\citealt{Cohen+2003}; however, the effect is small, with less than 2\% variation in relative magnitudes), and the $H$-band contains a water ice absorption band which could introduce similar variations.  Time-dependent absorption may therefore introduce temporary colour terms not evident in the all-night stack.  Therefore, we carefully inspected the time evolution of the observed zeropoints in all three bands.  A small amount (up to 0.3 mag) of total transmission variability is indeed observed during the first 20 minutes, after which the zeropoint in all three bands is nearly constant within uncertainties.   No significant variation is observed in the difference between zeropoints in different PAIRITEL bands, nor is any correlation observed between the overall zeropoint and the difference in zeropoints between two bands that would suggest chromatic variations in the transmission.  Furthermore, the zeropoint appears constant (within our uncertainties) after $\sim$1400~s (the SED is determined at 10000~s).  The MMTO cloud camera \footnote{skycam.mmto.arizona.edu} shows no evidence of significant cloud cover at any point during the night, and weather archives indicate warm and stable conditions during the observation.  Furthermore, in addition to PAIRITEL (Arizona), the Lick $J$-band (California) and MAGNUM (Hawaii) coeval measurements both give consistent results for the infrared magnitudes, giving additional confidence in our results; in particular both PAIRITEL and MAGNUM $JHK$ data sets show the putative extinction feature independently.   Therefore, we have no reason to believe that our SED is significantly affected by absorption features in Earth's atmosphere.

\begin{figure}
\centerline{
\includegraphics[scale=0.65,angle=0]{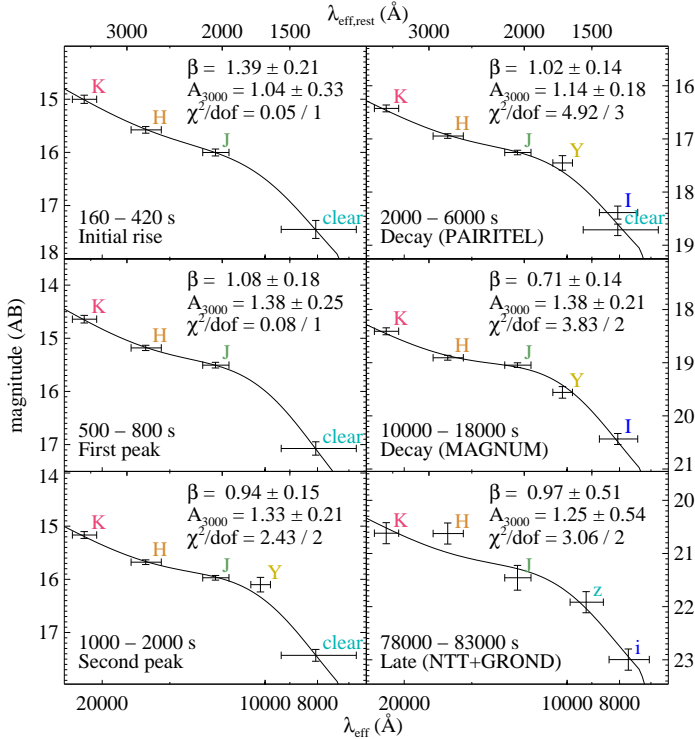}}
\caption{Time-dependent spectral energy distribution of GRB 071025 inferred after dividing the data into six different windows and re-fitting the flux parameters at each epoch using the light curve model.  The resulting SED is then fit for spectral index $\beta$ and extinction column $A_{3000}$ at each epoch individually using a Maiolino extinction profile.  The characteristic flattening between $J$- and $H$-bands is observed at every epoch (except at late times, when photometric errors are large) with no significant variation in its strength, increasing our confidence that it is a feature extrinsic to the GRB.}
\label{fig:multised}
\end{figure}

\textbf{Intrinsic deviation of the GRB spectrum from a power-law.}  We have assumed in our fits that the intrinsic spectrum of the GRB was a simple power-law, as generally predicted by synchrotron theory.  This assumption could, in principle, be violated.  However, the most natural deviation from a power-law SED that might be expected (a spectral break within the optical/IR band) would create downward curvature in the intrinsic SED and actually require additional dust to produce the upward inflection feature that is observed.   An SED modeled as the sum of two components (a steep power-law dominating $K-$band and a shallower power-law dominating $J$-band) would produce upward curvature, but cannot reproduce the sharpness of the observed feature unless the spectral index of the steep power-law is unrealistically red ($\beta>4$).  Additionally, it would be surprising that both components would rise and fall in synch with each other throughout the complex early evolution of the light curve, as is observed.  Indeed, evidence for Maiolino-like dust is observed at every epoch with no significant variation in its strength or wavelength (Figure \ref{fig:multised}; also \S\ref{sec:colevol}) with the exception of the final (GROND+NTT) SED, when the photometic uncertainties are too large to place any strong constraints on the extinction law.

\textbf{Absorption from a DLA host system or Lyman-$\alpha$ forest.}  If the host galaxy is at the maximum redshift of $z$=5.2, our mean-opacity model of the Lyman-$\alpha$ forest may significantly underestimate the impact of hydrogen absorption on the $I$-band.  To represent the most extreme possible case, we reran our dust models after adjusting the $I$-band flux upward by 20\% (the approximate maximum dimunition expected in the Kuiper $I$-band filter assuming 100\% opacity blueward of 7550\AA, the limit on any DLA imposed by the HIRES spectrum) at $z=5.2$.  Even in this case, the Maiolino dust profile is strongly preferred ($\chi^2$/dof = 2.5/2, versus 11.9/2 for SMC-like dust.)

\section{Discussion}

\subsection{Rise of the Forward Shock and Constraints on the Lorentz Factor}
\label{sec:rise}

The nearly achromatic first peak in the light curve shows the major hallmarks of the initial rise of the afterglow due to hydrodynamic deceleration of the fireball: a steep rise with no significant evidence of colour change across the peak.  Alternative possibilities can be generally ruled out:  for example, the transition of the synchrotron peak frequency (which would also produce a peak were it to occur after the initial deceleration) would rise slowly and exhibit a blue-to-red colour shift of $\Delta\beta = (p-1)/2 - (-1/3) = p/2 - 1/6$, completely incompatible with the observations of no colour change or even limited red-to-blue evolution at this time.  If the peak were due to dust destruction we would also expect significant colour change during the rise itself, which is not apparent in the data.  (We will examine the possibility of dust destruction in more detail in \S\ref{sec:colevol}).

Within the category of hydrodynamical effects, there are then three possibilities for the rise of the afterglow:  peak of the reverse shock, peak of the forward shock, or an off-axis jet.

We will consider the jet model \citep{Granot+2002,Granot2005} first.  In this case, the outflow is assumed to be strongly collimated with an observer located outside both the jet opening angle $\theta$ (observing angle $\theta_{\rm obs}$) and Lorentz cone $1/\Gamma$ (for a uniform jet; the theory can be suitably modified for a structured outflow: \citealt{Kumar+2003}).  As the jet decelerates, a peak in the light curve will be observed once the flow has decelerated sufficiently for the $1/\Gamma$ cone to expand past the observer line of sight; this model has shown reasonable success representing the rising light curves of e.g. XRF 080330 \citep{Guidorzi+2009} and GRB 080710 \citep{Kruhler+2009b}.  However, we are disinclined to favor this model on the grounds that it is expected to produce a very rapid post-break decay ($\alpha > 2$), which is not observed at late times ($\alpha_{2,f} = 1.27 \pm 0.04$).  This could be accounted for by associating the second component (which dominates the late-time decay) with an on-axis wide jet undergoing its initial rise (as in \citealt{Kruhler+2009b}), but this model is somewhat contrived in our case, requiring fine-tuning of the physical properties of the two jets to accommodate the large variation in their jetting times while still ensuring that they peak within a factor of $\sim$2 in time and flux.  Alternatively, refreshed shocks and continuous energy injection out to late times could also be invoked to explain the two-peaked structure and lack of late decay within this model.  Even in that case, another criticism of this model is that the isotropic energy release observed for this burst ($E_{\rm iso} = 6.5 \times 10^{53}$ erg) is not expected for a burst seen off-axis.

Next, we consider if the initial rise could be due to the reverse shock \citep{Sari+1999}.  This model is particularly attractive, as the overall light curve qualitatively looks impressively similar to the theoretical curve of \cite{Zhang+2003}: the first peak corresponds to the reverse shock and the second peak to the forward shock.  However, the initial rise is somewhat slower than expected from simple analytic models.  The assumed reverse shock rising index $\alpha_{1,r}$ depends on the assumed zero time $t_0$ (which was set to the trigger time in the above fits), but $t_0$ would need to be shifted back in time by an amount greatly in excess of the duration of the burst itself to match the predicted $t^{3p-5/2}$ predicted for the reverse shock rise in the slow-cooling case \citep{Kobayashi2000}.  The alternate fast-cooling case predicts a slower rise (too slow: $t^{13/16}$) and also a bluer spectrum than is preferred by our extinction modeling.  A wind model also requires fast-cooling and a blue spectrum, and an even slower rise ($t^{1/2}$).  Therefore a reverse shock is not our preferred paradigm either, though we are hesitant to rule it out on the basis that the known complexity of early afterglows and the failure of even late-time closure relations to properly predict the decay rate $\alpha$ \citepeg{Rykoff+2009} suggest that the quantitative details of light curve behavior may not be an especially reliable way to evaluate different models.

The most straightforward scenario for the initial rise is the formation of the forward shock as the burst ejecta decelerates into the surrounding medium \citepeg{Rees+1992}.  In this case $\alpha=-2$ for $\nu < \nu_c$, which is still somewhat too fast but still consistent with the data within $2\sigma$ if $t_0$ is moved backwards in time by about 30 s.  In this model, the second peak is presumably due to additional energy input from the central engine into the forward shock, perhaps in the form of a slow-moving shell that catches up at around 1 ks \citep{Rees+1998}.   This model is generally consistent with all available observations including the apparent rapid rise of the second component, though the observed significant (albeit minor) colour change is not predicted.  It could be due to the passage of a cooling break (though would imply $\nu > \nu_c$ initially and a too-steep $\alpha = -3$ during the rise) or another effect such as variation in the electron index $p$.

Interpreting this feature as a forward shock enables us to measure the initial Lorentz factor of the explosion.  Following e.g. \cite{Meszaros+2006} and \cite{Rykoff+2009}, this can be estimated from observable parameters via the following relationship:

$$\Gamma_0 = 2 \Gamma_{\rm dec} = 2 (\frac{3 E_{\rm iso}}{32 \pi n m_{\rm p} c^5 \eta t_{\rm pk,z}^3})^{1/8}$$
$$= 560 \frac{3 E_{\rm iso,52}}{\eta_{0.2} n_{0} t_{\rm pk,z,10}^3}$$

Here $E_{\rm iso,52}$ is the isotropic-equivalent energy release in units of $10^{52}$ erg, $\eta_{0.2}$ is the radiative efficiency in units of 0.2,  $n_0$ is the circumburst density in units of cm$^{-3}$, and $t_{\rm pk,z,10}$ is the afterglow peak time as observed at the burst redshift $z$ in units of 10 s.  For GRB~071025, using $E_{\rm iso} = 6.5 \times 10^{53}$ erg from our spectral model of the BAT data (at $z=5$), we derive:

$$\Gamma_0 \sim 206 \eta_{0.2}^{-1/8} n^{-1/8} $$  

Compared to direct pair-opacity lower limits inferred by the \Fermi LAT \citep{Abdo+2009a,Abdo+2009b}, this is a relatively low value of $\Gamma$.  However, it is fairly typical of afterglow-inferred values (100-1000, \citealt{Rykoff+2009}, \citealt{Molinari+2007}, \citealt{Oates+2009}, \citealt{Kruhler+2009a,Kruhler+2009b}).  This may indicate a difference in the types of populations probed by the two methods:  the intrinsic delay in optical follow-up can measure $\Gamma$ only for bursts for which the peak is quite late (low $\Gamma$), while high-energy photons themselves escape only if $\Gamma$ is large.  Hopefully, in the near future a joint \Swift-\Fermi burst with a luminous afterglow will allow both methods for estimation of the Lorentz factor to be compared.

\subsection{Colour Evolution: Limits on Dust Destruction}
\label{sec:colevol}

\begin{figure}
\centerline{
\includegraphics[scale=0.6,angle=0]{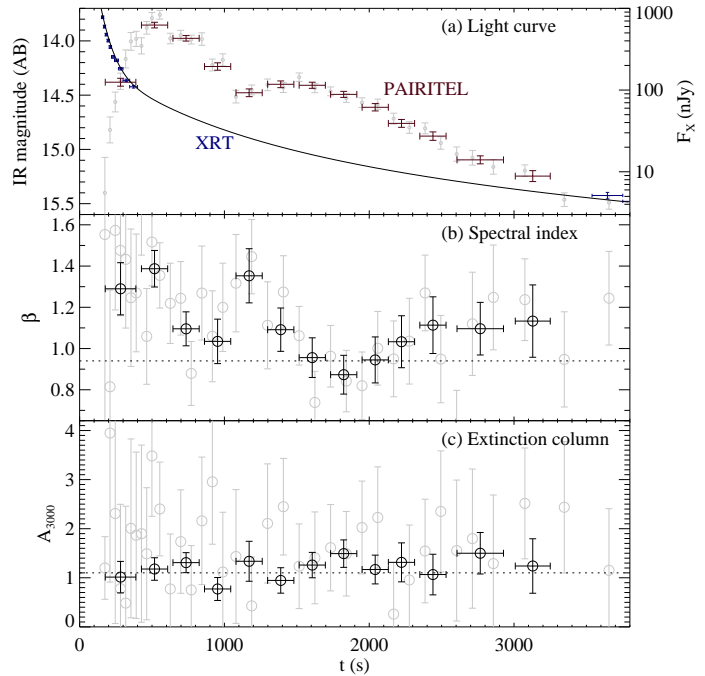}}
\caption{Models of colour evolution in the afterglow of GRB 071025.  (a) Infrared and X-ray light curves of GRB 071025 from PAIRITEL and the \Swift XRT showing the divergent behavior in the two bands at these times.  The early X-ray light curve is probably dominated by prompt emission, which is continuing in the BAT as well during this early decay phase.  (b) The infrared/optical spectral index $\beta$, as measured by a fit to PAIRITEL $JHK$ and RAPTOR unfiltered data.   Fixed extinction $A_{3000} = 1.1$ mag is assumed.  The SED is observed to redden significantly during the obervations.  Grey points indicate fits to PAIRITEL $JHK$ photometry only.  (c) The time-dependent extinction column $A_{3000}$ as measured by PAIRITEL and RAPTOR.  The spectral index $\beta$ is also free to vary in these fits.  No evidence for variation in the extinction column is observed, ruling out dust destruction after $\sim$150 s.}
\label{fig:colour1}
\end{figure}

Because of the need for $\nu_{\rm IR} < \nu_c$ to explain the slow rise, there is no explanation within the standard assumptions of afterglow theory for the colour change observed during the afterglow.  One possible solution would be to invoke a time-variable electron index $p$ at early times; a softening of the electron distribution during the complex early evolution would cause a corresponding softening of the afterglow emission.

Another intriguing possibility, however, is the photodestruction of dust along the GRB line of sight \citep{Waxman+2000,Draine+2002}.  While we have ruled out this model as being the predominant origin of the rise of the light curve based on the modest or absent colour change during the rising phase, it is still possible that it is occurring on a more subtle level.  Because our light curve model assumes any colour change is associated with a temporal break, it is not clear that such a change would be manifest in those models.  As a result, we have scrutinised the overall colour evolution of this GRB in significant additional detail to search for time-evolution in the extinction column $A_V$.

The large flat (grey) component of the Maiolino SN-type extinction law has the useful feature that the observed spectral slope of an SED measured over this region will closely match the intrinsic spectral slope even for a large extinction column, breaking the degeneracy between the intrinsic spectral index $\beta$ and amount of reddening $A_{3000}$.  At $z \sim 5$, the $J-H$ colour (where the extinction law is grey) is affected only by the intrinsic spectral index and is nearly independent of $A_{3000}$, while $H-K$ and $J-I$ are affected by both the intrinsic index and reddening.  This allows us to fit for $\beta$ and $A_{3000}$ independently with reasonable reliability, even with only a small number of points in the SED.

We have, therefore, undertaken time-variable extinction fits using simultaneous measurements by mosaicing the PAIRITEL $JHK$ data to temporally match the early-time RAPTOR points, which are a good approximation of the $I$-band (after a small adjustment of $-0.08$ mag: see \S \ref{sec:raptor} and Figure \ref{fig:lcurve}).  Dust models were fit to this four-point SED as in section \S \ref{sec:ext}.    Results are plotted in Figure \ref{fig:colour1}.

As in the case of the complete data set, a Maiolino dust model is significantly preferred, with no evidence of evolution.   In particular, the first mosaic (the only one contemporaneous with bright X-ray prompt emission, which is probably the dominant contributor to dust destruction: \citealt{Fruchter+2001}) gives a modest value of $A_{3000} = 1.03 \pm 0.31$, fully consistent with our measurement at 10000~s.  The corresponding 95\% confidence limit on the decrease in the extinction column is $\Delta A_{3000} < 0.54$ mag.   As we observe the $H$-band ($\lambda_{\rm eff} \sim 2770$\AA\ in the host frame) rising by at least 1.5 mag between our first REM and PAIRITEL observations and the peak, this clearly rules out dust destruction as the cause of the early peak, consistent with our conclusions of the chromatic light curve modeling in \S \ref{sec:rise}.    The entirety of the colour variation appears to be due to variation in the intrinsic spectrum.

The significant dust column, combined with the lack of variability even during the end of the prompt phase, places a limit on the proximity of this dust to the GRB.  The simulations of e.g. \cite{Perna+2003} suggest that for a bright GRB virtually all dust within about 10 pc of the GRB will be destroyed, and significant destruction will be observed even out to 100 pc.  While the exact constraints for this event will likely depend on detailed modeling of GRB~071025 specifically, this gives an approximate limit on the distance of the inferred absorbing dust column from the progenitor of at least $\gtrsim10-100$ pc.

\section{Conclusions}

GRB~071025 joins a growing list of gamma-ray bursts caught early enough in their evolution to observe the rise and peak of the optical afterglow.  Interpreting this as the initial rise of the forward shock, we estimate $\Gamma \sim 200$ for typical ISM densities.  The mild red-to-blue colour evolution of the afterglow appears to be due to unknown intrinsic properties of the forward shock, rather than dust destruction due to irradiation of the burst environment.   All of these properties are similar to those inferred from early-time observations of other GRB afterglows.

However, the extinction law we measure is nearly unique.  Most afterglows with well-characterised SEDs show little extinction \citep{Kann+2007b}, and events for which significant extinction has been observed have most commonly shown simple SMC-like profiles \citepeg{Kann+2006,Schady+2007}, characterised by significant curvature (strong wavelength dependence) but no spectral features.   More rarely, featureless or even grey light curves with no significant curvature have been inferred for some bursts \citepeg{Savaglio+2004,Stratta+2005,Chen+2006,Perley+2008a,Li+2008}, and recently a small number of events have been discovered containing the clear signature of the 2175-\AA\ bump present in the Milky Way and LMC \citep{Kruhler+2008,Prochaska+2009,Eliasdottir+2009}, though the details of these extinction curves show some differences from the average Milky Way ISM law.  But to our knowledge, no other GRB sightline has shown clear evidence of dust not well-fit either by a local extinction template or by a simple, featureless law.

A possible exception is $z$=6.3 GRB 050904.  For this GRB, an analysis by \cite{Stratta+2007} favored the supernova-type dust of \cite{Maiolino+2004} over standard SMC, Milky Way, and Calzetti models.  Taken together, these two bursts would represent compelling evidence of an association between the observed dust model and the chemical evolution of the universe itself: to date, these events are the only bursts at $z \gtrsim 4.5$ showing evidence for significant extinction (all other bursts for which useful constraints on the extinction law have been possible are at $z \lesssim 4$: \citealt{Kann+2007b}).\footnote{Unfortunately, the case of GRB 050904 is still ambiguous.  Numerous other papers have investigated the dust properties of this event \citep{Kann+2007,Gou+2007} and none of these other authors presented evidence favoring the Maiolino curve.  \cite{Liang+2009} have claimed detection of the 2175 \AA\ feature. Therefore, we downloaded the available data on this source \citep{Haislip+2006,Tagliaferri+2005,Kawai+2006,Boer+2006} and attempted to model the dust profile of this event using the same tools applied to GRB~071025, and found no evidence for a featured extinction curve.  Indeed, the data are fully consistent with no extinction at all: our extinction fits converged to a simple power-law with $\beta \sim 1.0$, in agreement with the comprehensive analyses of \cite{Kann+2007} and \cite{Gou+2007}.}

Strong chemical evolution of the dusty ISM is to be expected at $z \sim 5-6$: while most dust at low-to-moderate redshifts is thought to have been produced in AGB stars, during the first $\sim$1 Gyr following the Big Bang there had not yet been time for these stars to form in large numbers \citep{Morgan+2003}.  The cosmic age of $1.1-1.4$ Gyr (assuming standard cosmological parameters with $\Lambda$CDM) allowed by our photometric redshift suggests that SN-like dust\footnote{Although the extinction profile observed is an excellent match to the models of dust produced in supernovae provided by \cite{Maiolino+2004}, we note that this does not demonstrate conclusively that this dust was formed in the supernova explosion itself.  Alternatively, the dust could be formed in the ISM \citep{Draine2009} from refractory elements produced in SNe, from early carbon stars \citep{Sloan+2009}, or from an unknown pre-AGB formation mechanism.} is still the predominant source of obscuration in galaxies at this epoch and could provide important constraints on the evolution of the first galaxies and the production of early dust grains \citepeg{Dwek+2007}.  We note that the extinction column inferred from this galaxy is even larger than that inferred from the $z=6.2$ QSO ($A_{\rm 3000} = 0.4-0.8$ mag, \citealt{Maiolino+2004}), suggesting that even at this epoch, significant amounts of dust are present near sites of active star formation.  Alternatively, the unusual dust could be associated with the relatively nearby environment of the GRB only and not necessarily representative of the galaxy itself.  However, the survival of this late dust limits the distance from the progenitor to at least $10-100$ pc, suggesting that its association with the progenitor star-forming region cannot be too close.

The case of GRB~071025 is illustrative of the potential for early-time broadband photometry of GRBs to reveal the chemical history of the early universe \citep{Hartmann+2009}.  Well-characterised high redshift bursts are unfortunately rare (five years into the \Swift mission, only five other bursts to date have been confirmed to be at $z>5$), and high-redshift events showing significant dust columns are even rarer (with the exception of the controversial 050904, above, none of the other $z>5$ events show evidence for significant reddening: \citealt{Greiner+2009a,Ruiz-Velasco+2007,Tanvir+2009,Kann+2007b}.)  However, future GRB missions such as EXIST \citep{Grindlay+2009} are likely to produce a large advancement in our understanding of these events.  While designed to search for GRBs at the redshift extremes ($z>7$) and characterise these events spectroscopically, infrared photometry and spectroscopy acquired of the much more frequent moderate-redshift events ($z=4-7$) will place important constraints on the abundance and composition of dust during these early stages of cosmic evolution, when galaxies were in the active phase of assembly and the first generations of stars led to a rapid build up of the metal content of the universe.
 \\
 \\
 \\
 \\


PAIRITEL is operated by the Smithsonian Astrophysical Observatory (SAO) and was made possible by a grant from the Harvard University Milton Fund, a camera loan from the University of Virginia, and continued support of the SAO and UC Berkeley. The PAIRITEL project are further supported by NASA/\textit{Swift} Guest Investigator grant NNG06GH50G and NNX08AN84G.   RAPTOR/Thinking Telescopes project is supported by the Laboratory Directed Research and Development (LDRD) program at the LANL.  JXP is supported by NASA/Swift Guest Investigator grants NNX08AN90G and NNX09AO99G.  TK acknowledges support by the DFG cluster of excellence ``Origin and Structure of the Universe''.  Part of the funding for GROND (both hardware as well as personnel) was generously granted from the Leibniz-Prize (DFG grant HA 1850/28-1) to Prof. G. Hasinger (MPE).

We thank C. Melis at UCLA for acquiring the Lick infrared photometry.  We also thank D. A. Kann and the anonymous referee for useful comments and corrections on the manuscript.  The W. M. Keck Observatory is operated as a scientific partnership among the California Institute of Technology, the University of California, and the National Aeronautics and Space Administration (NASA). The Observatory was made possible by the generous financial support of the W. M. Keck Foundation.  We wish to extend special thanks to those of Hawaiian ancestry on whose sacred mountain we are priveleged to be guests.  This research has made use of the NASA/IPAC Extragalactic Database (NED) which is operated by the Jet Propulsion Laboratory, California Institute of Technology, under contract with the National Aeronautics and Space Administration.  We also acknowledge the hard work and efforts of the creators of other essential websites, in particular astrometry.net.


\begin{table*}
\begin{minipage}{80mm}
\centering
\caption{PAIRITEL Observations of GRB\,071025\label{tab:ptelphot}}
\begin{tabular}{rrrlll}
\hline
$t_{\rm start}$ & $t_{\rm end}$ & $t_{\rm exp}$ & $J$ & $H$ & $K_s$ \\
             s  &     s         &   s       & mag & mag & mag \\
\hline
$  162.3$ & $  186.9$ & $  23.4$ & $16.160 \pm 0.452$ & $15.933 \pm 0.768$ & $14.098 \pm 0.413$ \\
$  197.6$ & $  222.1$ & $  23.4$ & $15.448 \pm 0.153$ & $14.898 \pm 0.198$ & $13.751 \pm 0.195$ \\
$  233.8$ & $  258.2$ & $  23.4$ & $15.486 \pm 0.160$ & $14.589 \pm 0.143$ & $13.397 \pm 0.138$ \\
$  269.9$ & $  294.4$ & $  23.4$ & $15.300 \pm 0.141$ & $14.299 \pm 0.103$ & $13.259 \pm 0.119$ \\
$  306.1$ & $  330.6$ & $  23.4$ & $15.083 \pm 0.143$ & $14.072 \pm 0.124$ & $13.070 \pm 0.131$ \\
$  342.4$ & $  366.9$ & $  23.4$ & $14.814 \pm 0.126$ & $13.987 \pm 0.112$ & $12.905 \pm 0.136$ \\
$  378.6$ & $  403.1$ & $  23.4$ & $14.797 \pm 0.117$ & $13.956 \pm 0.110$ & $12.884 \pm 0.110$ \\
$  414.8$ & $  439.3$ & $  23.4$ & $15.058 \pm 0.129$ & $14.065 \pm 0.117$ & $12.854 \pm 0.112$ \\
$  451.1$ & $  475.5$ & $  23.4$ & $14.624 \pm 0.091$ & $13.817 \pm 0.086$ & $12.835 \pm 0.098$ \\
$  487.2$ & $  511.6$ & $  23.4$ & $14.669 \pm 0.084$ & $13.885 \pm 0.082$ & $12.606 \pm 0.079$ \\
$  523.3$ & $  584.0$ & $  46.8$ & $14.600 \pm 0.064$ & $13.774 \pm 0.064$ & $12.630 \pm 0.064$ \\
$  595.7$ & $  656.4$ & $  46.8$ & $14.801 \pm 0.083$ & $13.884 \pm 0.071$ & $12.919 \pm 0.080$ \\
$  668.1$ & $  728.8$ & $  46.8$ & $14.758 \pm 0.073$ & $13.913 \pm 0.069$ & $12.858 \pm 0.073$ \\
$  740.5$ & $  801.2$ & $  46.8$ & $14.677 \pm 0.059$ & $13.871 \pm 0.057$ & $13.007 \pm 0.073$ \\
$  813.9$ & $  874.6$ & $  46.8$ & $14.799 \pm 0.088$ & $13.977 \pm 0.083$ & $12.875 \pm 0.095$ \\
$  886.3$ & $  946.9$ & $  46.8$ & $14.942 \pm 0.083$ & $14.250 \pm 0.083$ & $13.136 \pm 0.090$ \\
$  958.6$ & $ 1019.4$ & $  46.8$ & $14.982 \pm 0.083$ & $14.100 \pm 0.076$ & $13.111 \pm 0.095$ \\
$ 1031.1$ & $ 1128.0$ & $  70.2$ & $15.359 \pm 0.099$ & $14.466 \pm 0.087$ & $13.417 \pm 0.094$ \\
$ 1139.8$ & $ 1236.7$ & $  70.2$ & $15.374 \pm 0.078$ & $14.355 \pm 0.061$ & $13.355 \pm 0.075$ \\
$ 1249.4$ & $ 1346.4$ & $  70.2$ & $15.161 \pm 0.080$ & $14.384 \pm 0.078$ & $13.330 \pm 0.090$ \\
$ 1358.1$ & $ 1456.0$ & $  70.2$ & $15.247 \pm 0.066$ & $14.443 \pm 0.063$ & $13.311 \pm 0.075$ \\
$ 1467.7$ & $ 1564.6$ & $  70.2$ & $15.086 \pm 0.052$ & $14.257 \pm 0.056$ & $13.296 \pm 0.068$ \\
$ 1576.3$ & $ 1673.0$ & $  70.2$ & $15.064 \pm 0.052$ & $14.351 \pm 0.061$ & $13.465 \pm 0.075$ \\
$ 1684.7$ & $ 1781.6$ & $  70.2$ & $15.176 \pm 0.052$ & $14.405 \pm 0.056$ & $13.435 \pm 0.073$ \\
$ 1793.3$ & $ 1890.2$ & $  70.2$ & $15.193 \pm 0.052$ & $14.452 \pm 0.054$ & $13.528 \pm 0.073$ \\
$ 1901.9$ & $ 1998.8$ & $  70.2$ & $15.227 \pm 0.054$ & $14.527 \pm 0.059$ & $13.554 \pm 0.082$ \\
$ 2010.5$ & $ 2107.3$ & $  70.2$ & $15.302 \pm 0.063$ & $14.564 \pm 0.066$ & $13.525 \pm 0.082$ \\
$ 2120.1$ & $ 2216.9$ & $  70.2$ & $15.434 \pm 0.066$ & $14.573 \pm 0.063$ & $13.739 \pm 0.094$ \\
$ 2228.6$ & $ 2325.6$ & $  70.2$ & $15.543 \pm 0.078$ & $14.702 \pm 0.066$ & $13.775 \pm 0.097$ \\
$ 2337.2$ & $ 2434.2$ & $  70.2$ & $15.637 \pm 0.066$ & $14.763 \pm 0.066$ & $13.713 \pm 0.087$ \\
$ 2445.9$ & $ 2542.8$ & $  70.2$ & $15.644 \pm 0.071$ & $14.927 \pm 0.078$ & $13.887 \pm 0.101$ \\
$ 2555.5$ & $ 2652.3$ & $  70.2$ & $15.602 \pm 0.082$ & $14.959 \pm 0.087$ & $14.111 \pm 0.125$ \\
$ 2664.0$ & $ 2760.9$ & $  70.2$ & $15.848 \pm 0.087$ & $15.039 \pm 0.087$ & $14.006 \pm 0.118$ \\
$ 2772.6$ & $ 2941.9$ & $ 117.0$ & $15.985 \pm 0.094$ & $15.100 \pm 0.084$ & $14.081 \pm 0.116$ \\
$ 2953.6$ & $ 3196.4$ & $ 163.8$ & $16.004 \pm 0.070$ & $15.211 \pm 0.072$ & $14.078 \pm 0.089$ \\
$ 3208.1$ & $ 3486.0$ & $ 187.2$ & $16.168 \pm 0.075$ & $15.455 \pm 0.084$ & $14.405 \pm 0.111$ \\
$ 3498.7$ & $ 3812.9$ & $ 210.6$ & $16.314 \pm 0.082$ & $15.420 \pm 0.075$ & $14.415 \pm 0.108$ \\
$ 9108.8$ & $12432.7$ & $2152.8$ & $17.680 \pm 0.154$ & $17.196 \pm 0.226$ & $16.074 \pm 0.228$ \\
$13132.0$ & $16637.0$ & $2269.8$ & $18.344 \pm 0.228$ & $17.462 \pm 0.245$ & $16.564 \pm 0.324$ \\
\hline
\end{tabular}
Time values are measured from the \Swift trigger (UT 2007 Oct 25 04:08:54).  Magnitudes are in the 2MASS (Vega) system and not corrected for Galactic extinction.
\end{minipage}
\end{table*}

\begin{table*}
\begin{minipage}{80mm}
\centering
\caption{Additional Photometry of GRB\,071025\label{tab:phot}}
\begin{tabular}{llccll}
\hline
Telescope & $t_{\rm mid}$ & Filter & $t_{\rm exp}$ & Mag. & Flux  \\
          &     s         &        & s             &      & $\mu$Jy \\
\hline
RAPTOR &   119.5 & {\rm clear} &   45.0 & $ > 16.94$ & $ <   608.9$ \\
RAPTOR &   290.4 & {\rm clear} &  200.0 & $17.187 \pm 0.159$ & $   485.9\pm    66.2$ \\
RAPTOR &   526.3 & {\rm clear} &  180.0 & $16.793 \pm 0.118$ & $   698.5\pm    71.9$ \\
RAPTOR &   739.6 & {\rm clear} &  180.0 & $16.792 \pm 0.111$ & $   699.2\pm    67.9$ \\
RAPTOR &   953.3 & {\rm clear} &  180.0 & $16.761 \pm 0.112$ & $   719.4\pm    70.5$ \\
RAPTOR &  1167.1 & {\rm clear} &  180.0 & $17.517 \pm 0.236$ & $   358.6\pm    70.1$ \\
RAPTOR &  1381.1 & {\rm clear} &  180.0 & $17.012 \pm 0.140$ & $   570.9\pm    69.1$ \\
RAPTOR &  1594.9 & {\rm clear} &  180.0 & $17.096 \pm 0.156$ & $   528.4\pm    70.7$ \\
RAPTOR &  1808.7 & {\rm clear} &  180.0 & $17.186 \pm 0.170$ & $   486.4\pm    70.5$ \\
RAPTOR &  2022.3 & {\rm clear} &  180.0 & $17.216 \pm 0.172$ & $   473.1\pm    69.3$ \\
RAPTOR &  2235.6 & {\rm clear} &  180.0 & $17.691 \pm 0.281$ & $   305.5\pm    69.7$ \\
RAPTOR &  2448.6 & {\rm clear} &  180.0 & $17.429 \pm 0.227$ & $   388.8\pm    73.4$ \\
RAPTOR &  2662.1 & {\rm clear} &  180.0 & $18.050 \pm 0.372$ & $   219.5\pm    63.7$ \\
RAPTOR &  2875.7 & {\rm clear} &  180.0 & $18.097 \pm 0.413$ & $   210.2\pm    66.5$ \\
RAPTOR &  3124.9 & {\rm clear} &  240.0 & $17.793 \pm 0.265$ & $   278.1\pm    60.2$ \\
Super-LOTIS &   134.5 & {\rm R}     &   50.0 & $ > 19.46$ & $ <   59.89$ \\
Super-LOTIS &   244.2 & {\rm R}     &  100.0 & $19.180 \pm 0.240$$^{\rm x}$ & $   77.51\pm   15.37$$^{\rm x}$ \\
Super-LOTIS &   478.7 & {\rm R}     &  300.0 & $19.700 \pm 0.240$$^{\rm x}$ & $   48.02\pm    9.52$$^{\rm x}$ \\
Super-LOTIS &   813.7 & {\rm R}     &  300.0 & $18.910 \pm 0.120$ & $   99.40\pm   10.40$ \\
Super-LOTIS &  1315.1 & {\rm R}     &  600.0 & $19.520 \pm 0.160$ & $   56.67\pm    7.77$ \\
Super-LOTIS &  1983.3 & {\rm R}     &  600.0 & $19.390 \pm 0.180$ & $   63.88\pm    9.76$ \\
REM &   470.0 & {\rm Y}     &   81.0 & $15.620 \pm 0.240$ & $  1257.2\pm   249.3$ \\
REM &  1281.0 & {\rm Y}     &  181.0 & $15.700 \pm 0.190$ & $  1167.9\pm   187.5$ \\
REM &  2653.0 & {\rm Y}     &  332.0 & $16.280 \pm 0.330$ & $   684.6\pm   179.4$ \\
REM &   377.0 & {\rm J}     &   91.0 & $15.220 \pm 0.200$ & $  1374.2\pm   231.2$ \\
REM &  1085.0 & {\rm J}     &  181.0 & $15.570 \pm 0.160$ & $   995.5\pm   136.4$ \\
REM &  2304.0 & {\rm J}     &  331.0 & $15.350 \pm 0.110$ & $  1219.1\pm   117.5$ \\
REM &   185.0 & {\rm H}     &   82.0 & $16.131 \pm 0.690$ & $   373.7\pm   175.8$ \\
REM &   666.0 & {\rm H}     &  181.0 & $13.890 \pm 0.050$ & $  2944.3\pm   132.5$ \\
REM &  1623.0 & {\rm H}     &  331.0 & $14.380 \pm 0.080$ & $  1874.9\pm   133.2$ \\
REM &  2795.0 & {\rm H}     &   81.0 & $ >14.90$ & $< 1161.4$ \\
REM &   275.0 & {\rm K}     &   82.0 & $13.430 \pm 0.130$ & $  2899.2\pm   327.2$ \\
REM &   890.0 & {\rm K}     &  212.0 & $12.900 \pm 0.070$ & $  4723.6\pm   294.9$ \\
Lick &  2714.0 & {\rm J}     &  210.0 & $15.817 \pm 0.030$ & $   792.9\pm    21.6$ \\
Lick &  2997.0 & {\rm J}     &  210.0 & $15.938 \pm 0.030$ & $   709.3\pm    19.3$ \\
Lick &  3279.0 & {\rm J}     &  210.0 & $16.135 \pm 0.030$ & $   591.6\pm    16.1$ \\
Lick &  3562.0 & {\rm J}     &  210.0 & $16.227 \pm 0.030$ & $   543.6\pm    14.8$ \\
Lick &  3846.0 & {\rm J}     &  210.0 & $16.378 \pm 0.030$ & $   473.0\pm    12.9$ \\
Lick &  4129.0 & {\rm J}     &  210.0 & $16.504 \pm 0.030$ & $   421.2\pm    11.5$ \\
Lick &  4413.0 & {\rm J}     &  210.0 & $16.558 \pm 0.030$ & $   400.7\pm    10.9$ \\
Lick &  4698.0 & {\rm J}     &  210.0 & $16.684 \pm 0.030$ & $   356.8\pm     9.7$ \\
Lick &  2714.0 & {\rm K}$^\prime$  &  210.0 & $14.143 \pm 0.100$$^{\rm x}$ & $  1503.4\pm   132.3$$^{\rm x}$ \\
Lick &  2997.0 & {\rm K}$^\prime$  &  210.0 & $14.321 \pm 0.100$$^{\rm x}$ & $  1276.1\pm   112.3$$^{\rm x}$ \\
Lick &  3279.0 & {\rm K}$^\prime$  &  210.0 & $14.384 \pm 0.100$$^{\rm x}$ & $  1204.1\pm   106.0$$^{\rm x}$ \\
Lick &  3562.0 & {\rm K}$^\prime$  &  210.0 & $14.518 \pm 0.100$$^{\rm x}$ & $  1064.3\pm    93.6$$^{\rm x}$ \\
Lick &  3846.0 & {\rm K}$^\prime$  &  210.0 & $14.583 \pm 0.100$$^{\rm x}$ & $  1002.5\pm    88.2$$^{\rm x}$ \\
Lick &  4129.0 & {\rm K}$^\prime$  &  210.0 & $14.650 \pm 0.100$$^{\rm x}$ & $   942.5\pm    82.9$$^{\rm x}$ \\
Lick &  4413.0 & {\rm K}$^\prime$  &  210.0 & $14.818 \pm 0.100$$^{\rm x}$ & $   807.4\pm    71.0$$^{\rm x}$ \\
Magnum & 10206.0 & {\rm J}     &  600.0 & $17.760 \pm 0.059$ & $   132.4\pm     7.0$ \\
Magnum & 10326.0 & {\rm R}     &  600.0 & $21.850 \pm 0.390$ & $   6.628\pm   2.000$ \\
Magnum & 11526.0 & {\rm Y}     &  300.0 & $18.408 \pm 0.183$ & $   96.43\pm   14.96$ \\
Magnum & 11526.0 & {\rm I}     &  600.0 & $19.880 \pm 0.140$ & $   30.49\pm    3.69$ \\
Magnum & 12846.0 & {\rm K}     &  480.0 & $16.491 \pm 0.086$ & $   172.9\pm    13.2$ \\
Magnum & 12846.0 & {\rm R}     &  600.0 & $ > 21.28$ & $ <   11.20$ \\
Magnum & 14106.0 & {\rm H}     &  540.0 & $17.597 \pm 0.082$ & $   96.87\pm    7.05$ \\
Magnum & 14106.0 & {\rm I}     &  600.0 & $20.220 \pm 0.200$ & $   22.29\pm    3.75$ \\
Magnum & 15366.0 & {\rm J}     &  540.0 & $18.320 \pm 0.103$ & $   79.08\pm    7.16$ \\
Magnum & 15366.0 & {\rm R}     &  600.0 & $ > 21.36$ & $ <   10.41$ \\
Magnum & 16506.0 & {\rm Y}     &  540.0 & $19.750 \pm 0.430$ & $   28.02\pm    9.16$ \\
Magnum & 16506.0 & {\rm I}     &  600.0 & $20.460 \pm 0.280$ & $   17.87\pm    4.06$ \\
Kuiper &  5098.5 & {\rm I}     & 1055.0 & $18.452 \pm 0.085$ & $   113.6\pm     8.6$ \\
Kuiper & 11260.0 & {\rm I}     & 2176.0 & $19.798 \pm 0.010$ & $   32.88\pm    0.30$ \\
Kuiper & 15382.0 & {\rm I}     & 3530.0 & $ > 19.65$ & $ <   37.68$ \\
Kuiper & 18584.5 & {\rm I}     & 2701.0 & $ > 19.23$ & $ <   55.48$ \\
Kuiper &  1824.8 & {\rm R}     &  261.7 & $19.290 \pm 0.040$ & $   70.05\pm    2.53$ \\
Kuiper &  2613.5 & {\rm R}     & 1225.9 & $19.790 \pm 0.040$ & $   44.20\pm    1.60$ \\
Kuiper &  3887.7 & {\rm R}     & 1259.3 & $20.300 \pm 0.060$ & $   27.63\pm    1.49$ \\
Kuiper &  8647.0 & {\rm R}     & 2940.0 & $ > 20.85$ & $ <   16.65$ \\
Kuiper &  6468.0 & {\rm V}     &  240.0 & $ > 21.4$ & $ <   12.73$ \\
NTT    & 81101.0 & {\rm J}     & 5104.0 & $20.780 \pm 0.270$$^{\rm x}$ & $   8.204\pm    1.806$$^{\rm x}$  \\
NTT    & 81672.0 & {\rm H}     & 4938.0 & $19.340 \pm 0.200$$^{\rm x}$ & $   19.45\pm    3.27$$^{\rm x}$ \\
NTT    & 82061.0 & {\rm K}     & 4960.0 & $18.780 \pm 0.200$$^{\rm x}$ & $   21.00\pm    3.53$$^{\rm x}$ \\
NTT    &  168032 & {\rm H}     & 2187.0 & $ > 19.80$ & $ <   12.73$ \\
GROND & 80505.0 & {\rm g}     &      -- & $> 23.2 $ & $  < 2.44$ \\
GROND & 80505.0 & {\rm r}     &      -- & $> 24.1 $ & $  < 0.997$ \\
GROND & 80505.0 & {\rm i}     &      -- & $23.140 \pm 0.270$$^{\rm x}$  & $   2.281\pm   0.502$$^{\rm x}$ \\
GROND & 80505.0 & {\rm z}     &      -- & $22.050 \pm 0.100$$^{\rm x}$  & $   6.163\pm   0.542$$^{\rm x}$  \\
GROND & 80533.0 & {\rm J}     &    2160 & $20.460 \pm 0.240$$^{\rm x}$ & $   11.02\pm    2.18$$^{\rm x}$ \\
GROND & 80533.0 & {\rm H}     &    2160 & $19.230 \pm 0.340$$^{\rm x}$ & $   21.53\pm    5.79$$^{\rm x}$  \\
GROND & 80533.0 & {\rm K}     &    2160 & $>18.13                    $ & $ < 38.3$  \\
\hline
\end{tabular}
Exposure mid-times are measured from the \emph{Swift} trigger (UT 2007 Oct 25 04:08:54).  $JHKYRI$ magnitudes are in the Vega system; $griz$ magnitudes are in the SDSS (approximately AB) system.  No Galactic extinction correction has been applied to magnitudes, but reported fluxes are corrected for $E_{B-V} = 0.07$ mag.  Limiting values are $3\sigma$.  Some $<3 \sigma$ detections are reported, as in many cases these are marginal detections that impose a useful constraint on the light curve or SED.  Points marked with an $^{\rm x}$ are not used in fitting.
\end{minipage}
\end{table*}

\begin{table*}
\begin{minipage}{80mm}
\centering
\caption{PAIRITEL $JHK_s$ Secondary Standards}
\begin{tabular}{llllllll}
\hline
$\alpha$ & $\delta$ & $J$ & $\sigma$ & $H$ & $\sigma$ & $K_s$ & $\sigma$ \\
    deg  &     deg  & mag &          & mag &          &       &          \\
\hline
355.107649 & 31.795298 & 11.681 & 0.012 & 11.445 & 0.011 & 11.418 & 0.007 \\
355.058508 & 31.790998 & 12.734 & 0.007 & 12.426 & 0.007 & 12.385 & 0.011 \\
355.037846 & 31.737404 & 13.095 & 0.007 & 12.717 & 0.007 & 12.627 & 0.010 \\
355.037722 & 31.708279 & 13.207 & 0.052 & 12.952 & 0.048 & 12.872 & 0.036 \\ 
355.134554 & 31.744791 & 13.704 & 0.031 & 13.393 & 0.029 & 13.343 & 0.024 \\
355.077212 & 31.725382 & 13.832 & 0.024 & 13.567 & 0.014 & 13.506 & 0.022 \\
355.105243 & 31.771235 & 14.137 & 0.007 & 13.762 & 0.007 & 13.709 & 0.015 \\
\hline
\end{tabular}
\label{tab:ircal}
Magnitudes are observed values, not corrected for Galactic extinction, and are reported in the 2MASS (Vega) system.
\end{minipage}
\end{table*}

\begin{table*}
\begin{minipage}{80mm}
\centering
\caption{Optical-IR Secondary Standards}
\begin{tabular}{llllllll}
\hline
$\alpha$ & $\delta$ & $R$ & $I$ & $Y$ & $J$ & $H$ & $K_s$ \\
     deg &  deg   & mag & mag & mag & mag & mag & mag \\
\hline
355.016646 & 31.770287 & 17.046 & 16.379  & 16.095 & 15.678 & 15.203 & 15.069 \\
355.025377 & 31.744127 & 13.699 & 12.469  & 11.816 & 11.297 & 10.683 & 10.476 \\
355.034260 & 31.771452 & 18.039 & 16.210  & 15.318 & 14.824 & 14.211 & 13.949 \\
355.037846 & 31.737404 & 14.231 & 13.684  & 13.418 & 13.095 & 12.717 & 12.627 \\
355.049875 & 31.745661 & 15.978 & 15.439  & 15.211 & 14.942 & 14.622 & 14.563 \\
355.058508 & 31.790998 & 13.740 & 13.229  & 12.990 & 12.734 & 12.426 & 12.385 \\
355.058815 & 31.780569 & 16.792 & 16.262  & 16.043 & 15.775 & 15.457 & 15.385 \\
355.066002 & 31.793428 & -      & 15.960  & 15.668 & 15.310 & 14.894 & 14.839 \\
355.068623 & 31.725666 & 17.911 & 16.674  & 17.607 & 16.794 & 16.042 & 15.371 \\
355.077212 & 31.725382 & 14.820 & 14.314  & 14.042 & 13.832 & 13.567 & 13.506 \\
355.105243 & 31.771235 & 15.358 & 14.765  & 14.452 & 14.137 & 13.762 & 13.709 \\
\hline
\end{tabular}
\label{tab:optcal}
Magnitudes are observed values, not corrected for Galactic extinction, and are reported in the Vega system.
\end{minipage}
\end{table*}

\end{document}